\documentclass[fleqn,10pt]{wlscirep}

\usepackage[utf8]{inputenc}
\usepackage[T1]{fontenc}

\usepackage{helvet}

\usepackage{sidecap}
\usepackage{float}
\usepackage[normalem]{ulem}
\usepackage{caption}

\DeclareCaptionLabelFormat{myformat}{\textbf{Fig}~#2}
\title{Wireless readout and control of cavity magnon-polaritons via photonic modes delocalised across a gap} 
\author[1,*]{Alban Joseph}
\author[2]{Ryan D McKenzie}
\author[1]{Amy Eden}
\author[1]{Morven Last}
\author[1]{Shannon Howard}
\author[1]{Alžbeta Oravcová}
\author[1]{Mawgan A Smith}
\author[2]{Robert L Stamps}
\author[1,\dag]{Rair Mac\^edo}
\affil[1]{James Watt School of Engineering, Electronics \& Nanoscale Engineering Division, University of Glasgow, Glasgow G12 8QQ, United Kingdom}
\affil[2]{Department of Physics and Astronomy, University of Manitoba, 30A Sifton Road, Winnipeg R3T 2N2, Manitoba, Canada}
\affil[*]{Alban.Joseph@glasgow.ac.uk}
\affil[$\dag$]{Rair.Macedo@glasgow.ac.uk}

\begin{abstract}

In cavity magnonics, photon–magnon hybridisation is typically detected by measuring the cavity response (i.e. measuring cavity resonance shifts through its reflection or transmission spectra). 
While a large body of work in the field has focused on three-dimensional cavities, a variety of two-dimensional resonators have also been explored as hosts for photonic modes due to their compatibility with on-chip integration.
Nevertheless, the readout approach remains largely unchanged. 
Here, we demonstrate remote excitation and readout of cavity magnon-polaritons using two spatially separated, wirelessly coupled ring resonators. Through inductive coupling, energy is transferred between the resonators, enabling the magnonic system to be excited by one resonator while its hybrid-mode spectrum is detected through measurements of the other. 
We demonstrate wireless readout of cavity magnon-polaritons at separations ranging from millimetres to centimetres, identifying their characteristic response in the remotely measured spectrum. This establishes wireless resonator coupling as a route to remote interrogation of hybrid photon–magnon systems.
\end{abstract}
\begin{document}

\flushbottom
\maketitle

\thispagestyle{empty}

\section*{Introduction}

The interaction between collective spin excitations---magnons---and electromagnetic modes confined within a microwave resonator can give rise to hybrid magnon-photon quasiparticles known as cavity magnon-polaritons\cite{macedo_CMPreview, rameshti_cavity_2022}.
When the magnon and photon modes are brought into resonance, coherent energy exchange between the two subsystems leads to their hybridisation, manifested experimentally by an avoided crossing and the emergence of distinct upper and lower polariton branches\cite{zhang14}. 
The strength of this interaction is governed by the magnon–photon coupling rate relative to the dissipation rates of the two modes,with the strong-coupling regime enabling coherent dynamics to be resolved within the system\cite{rameshti_cavity_2022}. 
These hybridised modes provide a means of transferring and encoding information between magnetic and electromagnetic degrees of freedom\cite{Huebl2013}, while their microwave response offers a direct route to the readout and control of the underlying magnetic excitations\cite{Joseph2024, joseph_dynamic_2026, Wolz2020}. 
Such capabilities have motivated the development of cavity magnonic systems for applications in microwave signal processing\cite{zhu2020,wang_lowloss_2021}, magnetic-field sensing\cite{Liu2025,Crescini2020}, and hybrid information-processing architectures\cite{li2020}, where controlled magnon–photon coupling can be exploited to engineer the properties of the microwave response\cite{yao2025_nonreciprocal,boventer2020_control, smith_exceptional_2026}.  

Three-dimensional microwave cavities coupled to yttrium iron garnet (YIG) spheres are often the systems of choice in cavity magnonics, owing to the low magnetic damping and narrow linewidth of YIG, together with the high quality factors achievable in three-dimensional cavity geometries. 
These characteristics enable strong magnon–photon coupling to be readily resolved through microwave spectroscopy. 
However, the relatively large physical dimensions of three-dimensional cavities can limit their integration into compact and scalable microwave architectures. 
In contrast, on-chip resonators offer significant advantages in terms of miniaturisation, integration, and scalability\cite{baity21_strong, hou2019strong, li2019strong,bhoi2017robust}. 
This integration can, however, come at the expense of resonator quality factor, $Q$, often  resulting in increased microwave losses and potentially weaker spectroscopic signatures of magnon–photon coupling\cite{castel2017thermal,shuai2025reconfigurable, macedo_electromagnetic_2021}.
Nonetheless, planar resonators trade $Q$ for integration, field confinement and geometric control of the coupling\cite{smith_exceptional_2026} and therefore, may be advantageous for practical implementations of on-chip cavity magnonic devices.

In addition to scalability, planar resonators offer additional functionality. 
For instance, ring resonators in particular display degenerate counter-propagating modes with weak backscattering at the feeds\cite{chang_microwave_2004} where coupling phase emerges as a design parameter\cite{avicena_coupling_2026} which can be designed through the resonator geometry.
Here, we will exploit another well-known functionality of these resonators: wireless energy transfer\cite{kurs_wireless_2007}. 
This is a well established mechanism in classical microwave theory that is applicable to two resonators at a distance where the electromagnetic fields between them allows for energy transfer across a wireless gap \cite{rehman_kazim_wireless_2023, kurs_wireless_2007, elnaggar_energy_2015, seetharaman_electromagnetic_2017}. 
By engineering a system where two ring resonators are strongly coupled through a wireless gap, we can use the coupling between them as a mechanism to readout and control of a magnon-photon hybridisation across a free-space gap, through a resonator with no magnetic element of its own.

In cavity magnonics, distributed coupling has been demonstrated where the magnonics system is typically read out through the resonator that hosts it, and remote magnon coupling has relied on a shared resonator\cite{li_coherent_2022} or on guided waves in transmission lines, coaxial cables and optical fibre\cite{rao_meterscale_2023, yang_anomalous_2024, joseph_dynamic_2026, xiong_magnon-photon_2024}. 
Here, on the other hand, we demonstrate that the magnon-photon hybridisation is read out through the second resonator, in both reflection and transmission, with no magnetic element on that chip.
This works because the separation between the rings redistributes the magnon coupling between the two hybridised branches, and the separation can be used to control the magnon coupling to each hybridised branch.  
% We do this by placing a YIG sphere on one planar ring resonator, whose coupling we first characterise as a function of position, $g \propto g_0|\cos n\theta|$; the resonator is then wirelessly coupled to a second, nominally identical ring across a variable gap $d$, with the inter-ring coupling $J$ tuned from well above the loaded linewidth to comparable with the magnon--photon coupling $g_0$.
The wireless link between two nominally identical rings is characterised as their separation $d$ is varied, which tunes the inter-ring coupling $J$ from well above the loaded linewidth to be comparable with the magnon--photon coupling. We then place a YIG sphere on one ring and characterise its coupling as a function of position, $g \propto g_0|\cos n\theta|$, before combining the two to read out and control the magnon through the unloaded ring. 
A result is that transmission provides a higher-contrast readout than reflection at large separation.

\section*{A wireless link between two rings}
\label{sec:link}

It is well known from microwave engineering that two resonators placed within each other's near field can exchange energy without any wired connection, through inductive coupling~\cite{kurs_wireless_2007}. A planar microstrip ring is a natural candidate for such coupling: as an open structure, part of the microwave magnetic field generated by the current on its track is not confined to the substrate, but fringes into the space above the board. A second ring placed within this field links part of its flux, so that the two rings are coupled through their mutual inductance. 
% This raises a simple question: can the same near-field coupling be used to address a magnon wirelessly, i.e. to drive and read out a YIG sphere sitting on one ring through a second ring that has no physical connection to it?
Before introducing any magnetic material, let us first look at this link on its own.

\begin{figure}
\centering
\includegraphics[width=1.0\linewidth]{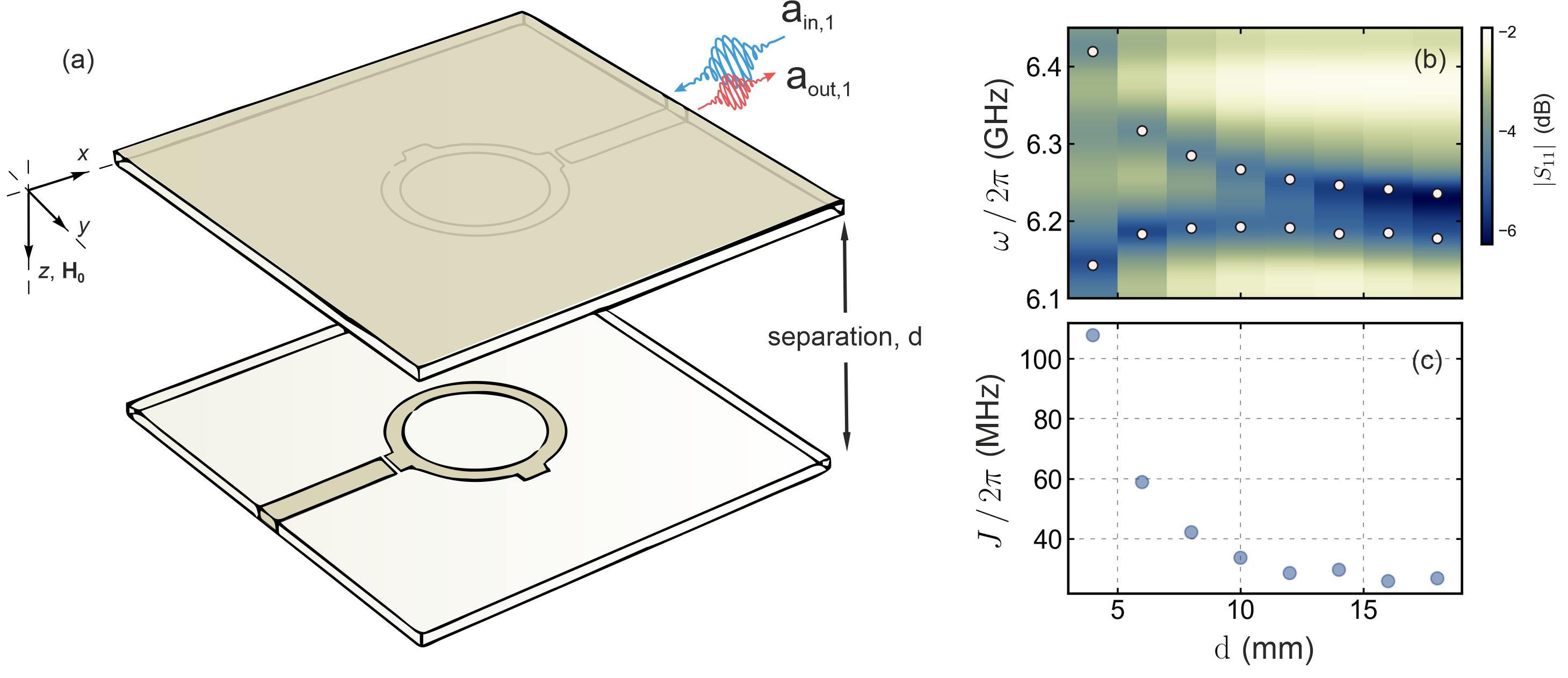}
\caption{\textbf{Wireless coupling between separated rings.} (a) Schematic of the stacked rings, separated by a distance $d$. $S_{11}$ is the reflection from ring~1 (upper, port~1), $a_{\rm out,1}/a_{\rm in,1}$. %, and $S_{12}$ is the transmission from port~2 to port~1, $a_{\rm out,1}/a_{\rm in,2}$. 
(b) Experimentally measured spectra $|S_{11}|$ versus $d$, with the extracted normal-mode frequencies overlaid as white circles. (c) Experimentally measured inter-ring coupling $J(d)$.}
\label{fig:fig2}
\end{figure}

% Let us now look at what happens when a second, nominally identical single-arm ring is mounted facing the first, with fixed lateral alignment and separation $d$ [Fig.~\ref{fig:fig2}(a)].
We mount two nominally identical single-arm rings facing one another, with fixed lateral alignment and separation $d$ as depicted in Fig.~\ref{fig:fig2}(a). We label the lower ring, connected to port~2, as ring~2, and the upper ring, connected to port~1, as ring~1. % The YIG sphere remains on the lower ring, held at $\theta = 0$ so that it couples maximally to the driven pattern. We label this lower ring, connected to port~2, as ring~2, and the upper ring, connected to port~1, as ring~1. 
Two measurements are then available: $S_{11} = a_{\rm out,1}/a_{\rm in,1}$, the reflection from ring~1, and $S_{12} = a_{\rm out,1}/a_{\rm in,2}$, the transmission from port~2 to port~1, which requires the signal to cross the gap between the rings. 

Treating each ring as a single resonant photon mode, $a_1$ and $a_2$, the link is described by
\begin{equation}
\frac{\mathcal{H}}{\hbar} = \tilde{\omega}_1 a_1^\dagger a_1 + \tilde{\omega}_2 a_2^\dagger a_2 + J a_1^\dagger a_2 + {\rm h.c.},
\label{eq:Hlink}
\end{equation}
where $\tilde{\omega}_{1(2)} = \omega_{1(2)} - i\kappa_{{\rm int},1(2)}$ are the complex resonance frequencies of the rings, including their internal loss, and $J$ is the inter-ring coupling, the rate at which the rings exchange energy through their shared field. 
% When the rings are brought together, the fringing field of each ring overlaps the other, so that the two rings exchange microwave energy coherently. The rate of this exchange is the inter-ring coupling $J$, set by the overlap of the two mode fields. 
Since the fringing field decays rapidly away from the board, $J$ falls steeply as the rings are separated, and $d$ therefore acts as a knob controlling how readily energy moves between them. Thus, what matters is how $J$ compares with the rate $\kappa_l$ at which each ring loses energy, through its port and internally.
% What matters is how $J$ compares with the two other rates in the system: the rate $\kappa_l$ at which each ring loses energy, and the rate $g_0$ at which ring~2 exchanges energy with the magnon.

When $J > \kappa_l$, energy passes between the rings faster than it is lost, and the two rings no longer respond as separate resonators. Instead, they form a pair of hybridised photonic modes, each of which extends over both rings and across the gap between them. 
% These modes no longer belong to either ring: each resides partly on both, so a signal launched into one ring appears in the other without any wired connection between them
A signal launched into either ring therefore sets up a field in both, and is carried to the other side without any wired connection. As $J$ falls below $\kappa_l$, energy is lost before it can be fully exchanged, and the two hybridised modes are no longer resolved. Energy nonetheless continues to cross the gap, so the link weakens rather than closes.

Independent reflection measurements on each bare ring give $\omega_2/2\pi = 6.23$~GHz, $\kappa_{l,2}/2\pi = 61.2$~MHz, $\kappa_{{\rm ext},2}/2\pi = 10.7$~MHz and $\kappa_{{\rm int},2}/2\pi = 50.5$~MHz ($Q_l = 51$, $Q_c = 291$, $Q_i = 62$) for ring~2, and $\omega_1/2\pi = 6.21$~GHz, $\kappa_{l,1}/2\pi = 62.97$~MHz, $\kappa_{{\rm ext},1}/2\pi = 15.8$~MHz and $\kappa_{{\rm int},1}/2\pi = 47.2$~MHz ($Q_l = 49$, $Q_c = 197$, $Q_i = 66$) for ring~1~\cite{probst_efficient_2015}, where each $\kappa$ term denotes the half-width at half-maximum of the corresponding resonance, such that $Q = \omega/2\kappa$. The two rings therefore have closely matched linewidths, with a mean loaded half-width $\bar{\kappa}_l/2\pi = 62.1$~MHz, and a residual detuning $\Delta = \omega_2 - \omega_1 = 2\pi \times 20.2$~MHz that we attribute to fabrication tolerance.

When the rings are brought together, their resonances split into the two hybridised modes, separated by $\Omega = \sqrt{\Delta^2 + 4J^2}$. Tracking these modes as a function of $d$ therefore gives $J(d)$ directly [Fig.~\ref{fig:fig2}(b)--(c)]. At the smallest separation, $d = 4$~mm, $J/2\pi = 107.8$~MHz, well above $\bar{\kappa}_l$: energy is exchanged between the rings faster than it is dissipated, and the two hybridised modes appear as clearly resolved resonances. At $d \approx 6$~mm, $J/2\pi = 58.9$~MHz, the system crosses over to $J \approx \bar{\kappa}_l$. Beyond $d \approx 10$~mm, the splitting falls below the resolution limit set by the overlap of two resonances of this width; $J$ can then no longer be obtained from $S_{11}$ alone, and values in this range are instead obtained from the dip positions in $S_{11}$, $S_{22}$ and the peak positions in $S_{12}$, $S_{21}$. The measured $J(d)$ is reproduced by finite-element simulations of the same geometry (Supplementary Fig.~\ref{fig:sup_fig_Jcompare}). Importantly, losing the resolved splitting does not mean energy transfer stops: a transmission band persists out to $d = 18$~mm, the largest separation measured.

The two rings therefore form a wireless link whose strength can be tuned continuously with separation, going from hybridised modes shared across the gap to weakly coupled rings that still exchange a fraction of their energy. This makes the link interesting not only in its own right, but also as a conduit. Any element placed on one ring, with no connection to the other, acts on modes that extend across the gap, and should therefore be visible from, and able to influence, the ring on the other side. In the following sections, we will test this idea with a magnon: we first load ring~2 with a YIG sphere and characterise its coupling, and then ask how the magnon, hosted on only one of the rings, can be read out and controlled through the other. 
Before doing so, we now characterise the behaviour of a cavity magnon-polariton where a YIG sphere is loaded onto a single ring resonator. 

\section*{System and model}
$ $

\begin{figure}[htb]
\centering
\includegraphics[width=0.8\linewidth]{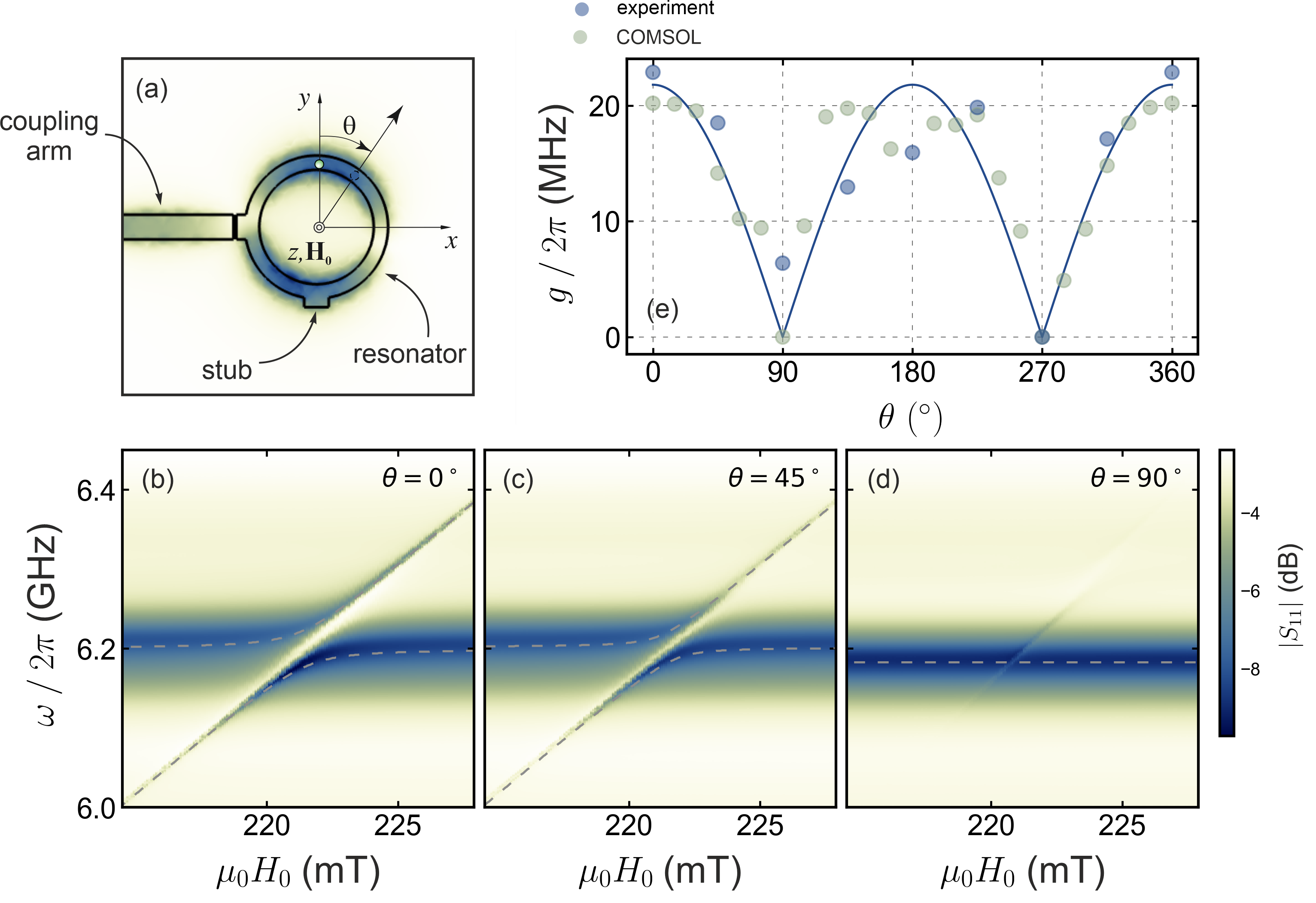}
\caption{\textbf{Resonator characterisation.} (a) Device geometry and simulated field distributions of the $n=1$ modes using COMSOL multiphysics. Experimental field-swept $|S_{11}|$ spectra at representative $\theta$, namely (b) $\theta = 0$, (c) $\theta = 45^\circ$, and (d) $\theta=90^\circ$, showing the position-dependent coupling behavior. (e) Experimentally extracted $g(\theta)$ (blue circles) and $|\cos \theta|$ fit (solid line), compared to COMSOL (green circles).}
\label{fig:fig1}
\end{figure}

%We now load ring~2 with a YIG sphere. 
To see how the magnon couples to the ring, we first need to look at the modes of a single ring in more detail. 
% We first characterise a single planar microstrip ring resonator and its coupling to a YIG sphere, as shown in Fig.~\ref{fig:fig1}. 
A ring supports travelling-wave resonances whenever its mean circumference is an integer number of guided wavelengths, $2\pi r=n\lambda_g$~\cite{chang_microwave_2004}, and we work with the $n=1$ resonance near $\omega_c/2\pi = 6.2$~GHz\cite{Hammerstad1980, pozar_microwave_2011}.  
At each of these frequencies there are two degenerate modes: the clockwise (CW) and counter-clockwise (CCW) travelling modes, here denoted $a_+$ and $a_-$ respectively. 
These can be understood as the combination of two orthogonal standing-wave patterns that can be mathematically written as $a_s = (a_+ + a_-)/\sqrt{2}$ and $a_c = (a_+ - a_-)/\sqrt{2}$ and whose magnetic-field profiles vary as $\sin(n\theta)$ and $\cos(n\theta)$, respectively, where $\theta$ denotes the angular position around the ring [Fig.~\ref{fig:fig1}(a)]. 
Since these two patterns are degenerate, the ring itself has no preference between them; instead, which one is excited is set by the position of the coupling arm. 
The arm couples capacitively to the ring voltage on the track where the two meet creating a voltage antinode, and hence a magnetic-field node, at the arm position. 
In other words, the standing wave is pinned to the coupling arm and moving the arm around the ring moves the excited pattern with it, along with it the regions of strong microwave magnetic field (which as we will later see are the regions of interest to excite magnons). 
For our geometry we excite the $\cos(\theta)$ pattern, while the orthogonal $\sin(\theta)$ pattern has a voltage node at the arm and is therefore dark to the exciting port.

% In practice, the picture of two independent patterns is not exact. The arm is a discontinuity on the track, so a wave travelling around the ring is partially reflected each time it passes it, scattering energy from the CW into the CCW mode and vice versa. We describe this by a backscattering rate $\beta = |\beta|e^{i\varphi_\beta}$. Its real part shifts the two standing-wave patterns in opposite directions, lifting their degeneracy, while its imaginary part mixes them; together, the eigenmodes are split by $2|\beta|$. The practical consequence is that the nominally dark pattern is no longer fully dark: part of the excitation launched by the port leaks into it.

% the spatial phase of the travelling modes produces position-dependent coupling, and the standing-wave decomposition makes the bright/dark behaviour transparent
A YIG sphere of diameter 0.25~mm is mounted on a ceramic rod and positioned on the ring at angular position $\theta$, as shown in Fig.~\ref{fig:fig1}(a). 
Its Kittel mode, with frequency $\omega_0 = \gamma \mu_0 H_0$ (where $\gamma/2\pi = 28$~GHz/T for YIG), is tuned by an external static magnetic field, $\mathbf{H_0}$, and couples to the local microwave magnetic field of the ring. 
Projecting this microwave field onto the two standing-wave patterns (as shown in Supplementary Note~B), the magnon couples to them with strengths $g_c = -ig_0\cos(\theta)$ and $g_s = -g_0\sin(\theta)$, respectively. The sphere therefore couples to the driven pattern with strength $g_0|\cos(\theta)|$, and when placed at its node, the magnon couples only to the dark pattern and leaves no signature in the reflection spectrum. 

We note that our resonators contain an additional feature: \emph{`stubs'}. 
In practice, the stubs couple the CW and CCW modes through a backscattering rate $\beta = |\beta|e^{i\varphi_\beta}$. 
The coherent coupling can lift the degeneracy, splitting the eigenmodes by $2|\beta|$, while its imaginary part mixes the two standing-wave patterns, transferring part of the excitation into the nominally dark one.
We have experimentally characterized this effect using a two-arm ring otherwise identical in geometry and we estimate $|\beta|/2\pi \approx 18$~MHz. The resulting splitting, $2|\beta|/2\pi \approx 36$~MHz, lies well within the loaded linewidth, $2\kappa_l/2\pi \approx 122$~MHz, and in reflection it appears only as a slight broadening of the resonance. The CW and CCW modes are therefore effectively degenerate on the scale of the cavity linewidth, and the reflection response is determined primarily by how the ring is driven.
Therefore, while backscattering may yield interesting new physics---particularly physics similar to that explored in Ref[\cite{smith_exceptional_2026}]---it does not significantly affect the qualitative behaviours of interest in this study. 
Nonetheless, experimental and theoretical expansion of this work using backscattering is given in the Supplemental Notes for completeness. 

We can then write the full Hamiltonian for the single ring system in the standing wave basis as
\begin{equation}
\frac{\mathcal{H}}{\hbar} = \tilde{\omega}_a \left( a_c^{\dagger}a_c 
+ a_s^{\dagger}a_s \right)
+ \widetilde{\omega}_0\, m^{\dagger}m
+ \biggr[g_c a_c^{\dagger} m 
+ g_s a_s^{\dagger} m + {\rm H.c.}\biggr],
\label{eq:Hfull}
\end{equation}
where $\widetilde{\omega}_a = \omega_a - i\kappa_{\rm int}$ and $\tilde{\omega}_0 = \omega_0 - i\kappa_m$ are complex frequencies that include the intrinsic damping of the ring ($\kappa_{\rm int}$) and magnon modes ($\kappa_m$), respectively. The sine and cosine standing waves are degenerate in the absence of backscattering.

% The reflection response of the bare single-arm ring at its $n=1$ resonance was analysed, yielding a resonance frequency $\omega_c/2\pi = 6.23$~GHz and a loaded half-width $\kappa_l/2\pi = 61.2$~MHz. The loaded linewidth decomposes into external and internal contributions of $\kappa_{\rm ext}/2\pi = 10.7$~MHz and $\kappa_{\rm int}/2\pi = 50.5$~MHz, corresponding to $Q_l = 51$, $Q_c = 291$, and $Q_i = 62$~\cite{probst_efficient_2015}, where $\kappa$ denotes the half-width at half-maximum of the corresponding resonance, such that $Q = \omega/2\kappa$. 
% % The relatively low intrinsic quality factor, $Q_i = 62$, indicates an additional radiative loss channel arising from weakly confined fields that extend above the board and will mediate coupling between the separated rings. 
%The complex coupling $\beta$ describes coherent backscattering between the CW and CCW modes, while $\beta_d$ accounts for a dissipative contribution to this coupling.

% $\omega_c/2\pi = 6.2308$~GHz, $\kappa_l/2\pi = 61.227$~MHz, $\kappa_{\rm ext}/2\pi = 10.703$~MHz and $\kappa_{\rm int}/2\pi = 50.524$~MHz, corresponding to $Q_l = 51$, $Q_c = 291$, and $Q_i = 62$. $\kappa_m/2\pi = 3.9$~MHz.
% $\omega_c/2\pi = 6.2106$~GHz, $\kappa_l/2\pi = 62.966$~MHz, $\kappa_{\rm ext}/2\pi = 15.799$~MHz and $\kappa_{\rm int}/2\pi = 47.168$~MHz ($Q_l = 49$, $Q_c = 197$, $Q_i = 66$).

%\begin{SCfigure}[][ht]
%\end{SCfigure}

%\subsection*{Position dependence of the magnon--photon coupling}
%\label{sec:gtheta}
%$ $

% We now turn to the position dependence of the magnon--photon coupling. 
To connect the Hamiltonian in Eq.~\eqref{eq:Hfull} to the experimentally measured spectra, we employ input--output theory~\cite{GardinerCollett,GardinerZollerBook,Clerk2010}. 
%Since the backscattering splitting lies well within the cavity linewidth, we set $\beta = \beta_d = 0$ here for clarity; the full expressions, including backscattering, are given in Supplementary Note~B. % The feed then couples only to the driven $\cos(n\theta)$ pattern, at a rate $\kappa_{\rm ext}$, while the $\sin(n\theta)$ pattern is dark and decays only through its internal loss, $\kappa_{\rm int}$. 
The reflection coefficient takes the form
\begin{equation}
S_{11} =1-\cfrac{i2\kappa_{\rm ext}}{ \omega- z_{c} - \cfrac{g_0^2 \cos^2 (\theta)}{\omega-z_m} - \cfrac{g_0^4 \sin^2 (2\theta)\, \chi_m^2/4}{\omega- z_s - \cfrac{g_0^2 \sin^2 (\theta)}{\omega-z_m}}},
\label{eq:S11}
\end{equation}
where $z_{c(s)} = \tilde{\omega}_a - i\kappa_{\rm ext}$ are the complex frequencies of the driven ($z_c$) and dark ($z_s$) patterns, $z_m = \tilde{\omega}_0$, and $\chi_m = (\omega - z_m)^{-1}$ is the magnon response.  
The final term in the denominator of Eq.~\eqref{eq:S11} describes an indirect pathway: the driven pattern excites the magnon, which in turn exchanges energy with the dark pattern before returning to the driven one. 
This higher order scattering process has little impact on the reflection spectrum. 
At the antinodes of the driven ($z_c$) pattern, $\theta = 0\ \text{and}\ \pi$, the indirect term vanishes and the magnon couples to the driven pattern with the full strength $g_0$. At the nodes, $\theta = \pi/2 \ \text{and}\ 3\pi/2$, both magnon terms vanish and $S_{11}$ reduces to the bare cavity response, so the magnon leaves no signature in reflection, even though it remains coupled to the dark pattern, which has a finite magnetic field at the same position.
% Its numerator requires the magnon to couple to both patterns at once ($\sin^2(2n\theta)$ dependance), so this pathway is largest at $\theta = \pi/(4n)$ and vanishes at both the antinodes and the nodes of the driven pattern. Its denominator, $D_{sm}^{-1}$, is the response of the dark pattern, which decays only at $\kappa_{\rm int}$ and is itself dressed by the magnon through $g_0^2\sin^2(n\theta)/(\omega - z_m)$.

% When the sphere sits at a node of the driven pattern, $\theta = \pm\pi/(2n)$, $g_c$ vanishes and the magnon leaves no signature in $S_{11}$, even though it remains coupled to the dark pattern, which has a finite magnetic field at the same position.
% At these positions the dark pattern has a magnetic-field maximum, so the magnon remains fully coupled to it; the coupling has simply moved to a mode that the port cannot see.

We translate the YIG sphere around the ring track and, at each position, sweep the static field through the cavity resonance. The resulting field-swept reflection spectra [Fig.~\ref{fig:fig1}(b)--(d)] show a clear avoided crossing between the magnon mode and the ring resonance, whose size depends strongly on $\theta$. %Fitting each spectrum to Eq.~\eqref{eq:S11} yields the coupling to the driven pattern, $|g_c(\theta)|$, shown in Fig.~\ref{fig:fig1}(e). [
The reflection minima predicted by Eq.~\eqref{eq:S11} are overlaid as dashed lines in Fig.~\ref{fig:fig1}(b)--(d). The coupling reaches its maximum, $g_0/2\pi = 21.8$~MHz, at the antinodes, $\theta = 0\ \text{and}\ \pi$, and follows the expected $|\cos(\theta)|$ dependence around the track [Fig.~\ref{fig:fig1}(e)]. The magnon half-width is $\kappa_m/2\pi = 3.9$~MHz, giving a cooperativity $C = g_0^2/(\kappa_l\kappa_m) \approx 2$ at the antinode~\cite{Huebl2013}.
% $C = \frac{g^2}{\kappa_l\kappa_m}=21.8^2/(62.966*3.9)$
% at the antinodes, $\theta = m\pi/n$ for integer $m$

% % With the present choice of angular origin, the measured coupling therefore follows the spatial magnetic-field profile of the mode addressed by the feed. When the YIG sphere is positioned at a node of this mode, the measured magnon--cavity coupling vanishes. 
% % Importantly, this vanishing coupling does not imply that the YIG is decoupled from the ring. At these positions, the orthogonal $\sin(n\theta)$ mode has a finite magnetic field and therefore remains coupled to the magnon. The coupling is consequently transferred from the port-bright mode to the port-dark mode, becoming accessible in the two-arm transmission measurement. %

\subsection*{Magnon control of the inter-ring transmission}

\begin{figure}[hbt]
\centering
\includegraphics[width=\linewidth]{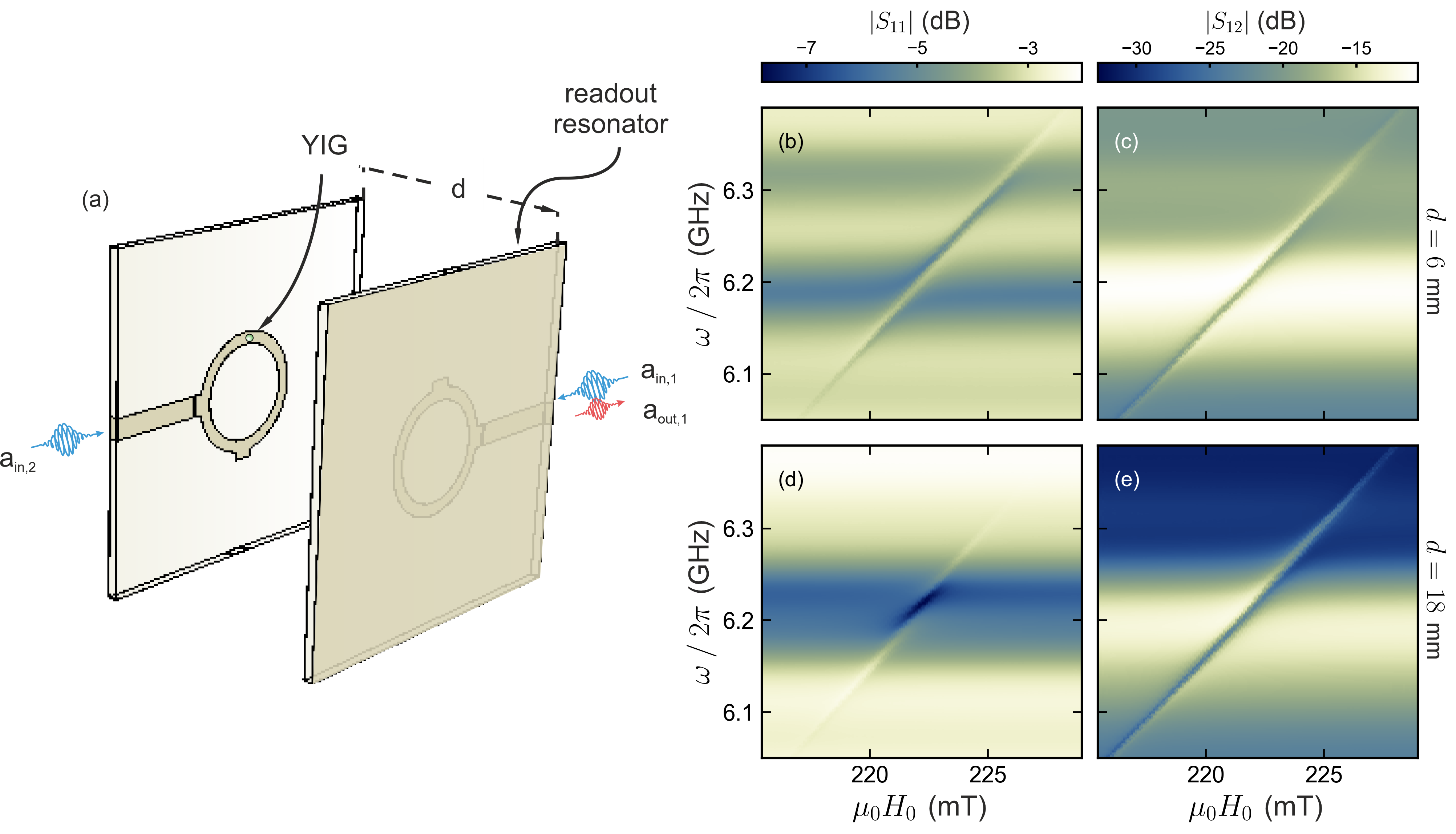}
\caption{\textbf{Magnon hybridisation across the gap.} (a) Measurement configuration: the YIG sphere sits at $\theta = 0$ on ring~2; ring~1, the readout resonator, carries no magnetic material. (b)--(e) Field-swept experimental spectra at $d = 6$~mm (b),(c) and $d = 18$~mm (d),(e). Left: reflection at port~1, $|S_{11}| = |a_{\rm out,1}/a_{\rm in,1}|$. Right: transmission between the rings, $|S_{12}| = |a_{\rm out,1}/a_{\rm in,2}|$.}
\label{fig:fig3}
\end{figure}

We now bring the two parts together. The YIG sphere is held at $\theta = 0$ on ring~2, where it couples maximally to the driven $\cos\theta$ pattern and not directly to the dark $\sin\theta$ pattern, and ring~1 is placed above it, as in Fig.~\ref{fig:fig3}(a). Since backscattering mixes the two patterns only weakly on the scale of the loaded linewidth, the dark patterns can be neglected, and the system is described by adding the magnon to Eq.~\eqref{eq:Hlink},
% As established above, backscattering mixes these patterns only weakly on the scale of the loaded linewidth, so we neglect it here; the dark patterns can then be excluded, leaving the two driven cavity modes and the magnon [Fig.~\ref{fig:fig2}(a)]. The resulting three-mode Hamiltonian is described by adding the magnon to Eq.~\eqref{eq:Hlink},
\begin{equation}
\frac{\mathcal{H}}{\hbar} = \tilde{\omega}_1 a_1^\dagger a_1 + \tilde{\omega}_2 a_2^\dagger a_2 + \tilde{\omega}_0 m^\dagger m
+ \biggr[ J a_1^\dagger a_2 -i g_0 a_2^\dagger m + {\rm H.c.} \biggr],
\label{eq:H2ring}
\end{equation}
where $a_{1(2)}$ is the driven pattern of ring~1(2) and $\tilde{\omega}_{1(2)} = \omega_{1(2)} - i\kappa_{{\rm int},1(2)}$. \textcolor{red}{} 
% Ring~2 is the ring characterised above, while independent measurements on ring~1 give $\omega_1/2\pi = 6.21$~GHz, $\kappa_{l,1}/2\pi = 62.97$~MHz, $\kappa_{{\rm ext},1}/2\pi = 15.8$~MHz and $\kappa_{{\rm int},1}/2\pi = 47.2$~MHz ($Q_l = 49$, $Q_c = 197$, $Q_i = 66$). The two rings therefore have closely matched linewidths, with a mean loaded half-width $\bar{\kappa}_l/2\pi = 62.1$~MHz, and a residual detuning $\Delta = \omega_2 - \omega_1 = 2\pi \times 20.2$~MHz that we attribute to fabrication tolerance. 
The magnon and ring~1 never interact directly. Any information about the magnon that reaches port~1 must pass through ring~2, which acts as the conduit between the magnon and the readout resonator.

Following the same input--output procedure, with $z_{1(2)} = \tilde{\omega}_{1(2)} - i\kappa_{{\rm ext},1(2)}$ and $z_m = \tilde{\omega}_0$, the reflection coefficient at port~1 is
\begin{equation}
S_{11} = 1 - \cfrac{i2\kappa_{{\rm ext},1}}{\omega - z_1 - \cfrac{J^2}{\omega - z_2 - \cfrac{g_0^2}{\omega - z_m}}},
\label{eq:S2ring_refl}
\end{equation}
and the transmission through port 2 is
\begin{equation}
S_{21} = S_{12} = \frac{i2\sqrt{\kappa_{{\rm ext},1}\kappa_{{\rm ext},2}}\, J\,(\omega - z_m)}
{(\omega - z_1)\left[(\omega - z_2)(\omega - z_m) - g_0^2\right] - J^2(\omega - z_m)}.
\label{eq:S2ring_trans}
\end{equation}
The nested structure of Eq.~\eqref{eq:S2ring_refl} follows the path of the energy. 
The port drives ring~1, ring~1 passes energy to ring~2 at rate $J$, and ring~2 exchanges it with the magnon at rate $g_0$ before it returns. The magnon is therefore read out wirelessly, appearing in the reflection from ring~1 only through the inter-ring coupling. In transmission, the signal launched into ring~2 must cross to ring~1 to reach port~1, so $S_{12} \propto J$. 

Fig.~\ref{fig:fig3}(b)--(e) shows the reflection at port~1 and the transmission between the ports at $d = 6$ and 18~mm; $S_{22}$, the reflection of the ring carrying the sphere, is given in Supplemental Note~C.

The magnon appears clearly in the reflection from ring~1 [Fig.~\ref{fig:fig3}(b),(d)], even though ring~1 carries no magnetic material. Microwave energy entering at port~1 is passed to ring~2 through the inter-ring coupling, where it exchanges energy with the magnon before returning; ring~1 therefore reads out the magnon across the gap without any magnetic element of its own.

For degenerate resonators, the hybridised modes are equal mixtures of the bare resonator modes. 
Rewriting Eq.~\eqref{eq:H2ring} in the basis of the hybridised ring modes, the system reduces to the same three-mode structure as in Ref.~\cite{smith_exceptional_2026},
where the magnon, although hosted only on ring~2, couples to both with the same strength, $g_0/\sqrt{2}$, at every separation. 
With a detuning of the resonators, as in the experiment, this is no longer the case. As a consequence, the couplings between the magnon and the hybridised resonator modes will depend on the separation of the resonators. 
%How strongly the magnon appears in each feature of the spectrum is set by the hybridised ring modes. 
The magnon couples only to ring~2, so it couples to each hybridised mode in proportion to how much of that mode resides in ring~2. At small separations, where the inter-ring coupling is much stronger than the detuning, the two modes are shared almost evenly between the rings, and the magnon couples to both with similar strength. As the rings are drawn apart and $J$ falls, each mode moves towards its own ring: since ring~2 is the lower in frequency, the lower mode comes to reside increasingly in ring~2 and the upper mode in ring~1. The magnon therefore couples more strongly to the lower mode and more weakly to the upper one.

% {\color[RGB]{201,57,57}Experimentally, the separation changes how each mode is shared between the two rings, and hence how quickly the magnon exchanges energy with each branch of the link. The static field can then select which branch the magnon is brought into resonance with, this would turn the separation and the field into two independent controls: $d$ sets how the magnon's coupling is divided between the branches, and $H_0$ chooses which one it acts on.}
This is what we observe: in the reflection of ring~1, the coupling to the lower mode rises from $17.0$ to $18.5$~MHz between $d = 4$ and $12$~mm, while that to the upper mode falls from $13.5$ to $12.0$~MHz. Separating the rings therefore does not simply weaken the magnon's interaction with the system; it redistributes it, concentrating the coupling into the mode that resides mostly on ring~2. The separation also sets how the hybridised modes appear in the spectrum, and hence how the magnon crosses them. At $d = 6$~mm, $J \approx 2.7g_0$ and the two modes are separated by roughly twice their linewidth, so the magnon meets each of them in turn as the field is swept, producing two avoided crossings [Fig.~\ref{fig:fig3}(b)]. At $d = 18$~mm, $J \approx 1.2g_0$ and the modes lie well within each other's linewidth; they are no longer individually resolved, the upper crossing (already unresolved beyond $d \approx 10$~mm) is lost, and the interaction is concentrated into a single region [Fig.~\ref{fig:fig3}(d)].

The transmission spectrum carries the same information about the magnon across the gap [Fig.~\ref{fig:fig3}(c),(e)]. Here the magnon meets a single transmission band at every separation, and its coupling to it rises from $15.5$ to $19.5$~MHz between 4 and 12~mm, following the same trend as the lower mode in reflection: as the rings separate, the magnon's interaction with the transmitted signal strengthens rather than weakens.
% {\color[RGB]{201,57,57} Why does transmission show a single band rather than the two hybridised modes seen in reflection? This band is the wireless link itself, so understanding which modes carry it, and how, tells us what the magnon is actually acting on when it modulates the transmitted signal.} 
% At large separations, the magnon is in fact far more visible in transmission than in reflection. In $S_{12}$, the magnon line remains clearly traceable across the whole field sweep, with a distinct avoided crossing and antimode where it meets the cavity band [Fig.~\ref{fig:fig3}(e)], whereas in $S_{11}$ the feature fades into the cavity band [Fig.~\ref{fig:fig3}(d)]. This contrast follows directly from how the energy flows in each measurement. Ring~1 is strongly undercoupled, so most of the power incident on port~1 is reflected straight back, regardless of what happens in ring~2; the magnon modifies only the small fraction that reaches ring~2 and returns. In transmission there is no such background: every photon reaching port~1 from port~2 must first pass through ring~2, where the magnon sits, so the magnon acts on the entire transmitted signal. The transmission is therefore the natural channel through which the magnon controls the wireless link, which we examine in detail below. 

Furthermore, we note there is strong asymmetry in the amplitude and linewidths of the hybridised cavity modes. This stems from the lossy nature of the coupling between the resonators. The low-frequency hybridised mode visible in the transmission spectrum in Fig. \ref{fig:fig3} (c) is stronger and sharper than the high-frequency mode visible above $6.3$ GHz. As elaborated on in Supplemental Notes C and D, this behaviour can be modelled by including a dissipative component in the inter-ring coupling not included in the simplified model given in Eq. (\ref{eq:H2ring}). 

The two measurements offer complementary ways of accessing the magnon. In reflection, only ring~1 needs to be connected and the magnon is read out through a resonator that carries no magnetic material, and ring~2 requires, in principle, no wired connection at all. 
This is the natural configuration when the magnetic element cannot, or should not, be wired directly (at the cost of contrast). %: ring~1 is strongly undercoupled, so most of the incident power is reflected straight back regardless of the magnon, which modifies only the small fraction that reaches ring~2 and returns. At large separations the magnon feature in $S_{11}$ therefore fades into the cavity band [Fig.~\ref{fig:fig3}(d)].
In transmission, the two rings already form a wireless link without the magnon, and the magnon modifies the signal crossing it. Since every photon crossing the link must pass through ring~2, where the magnon sits, the magnon acts on the entire transmitted signal.
Its line remains clearly traceable across the whole field sweep at every separation, with a distinct avoided crossing and antimode where it meets the cavity band [Fig.~\ref{fig:fig3}(e)]. The transmission is therefore the natural channel through which the magnon controls the wireless link, which we examine in detail below.

% % θ sweep: use 12 mm and 6 mm. At 12 mm, J ≈ 32 MHz and g₀ ≈ 20–27 MHz are comparable, so varying g through θ reshapes the eigenmodes, and it's where the modulation depth has saturated. Do 6 mm as the contrast case, where J ≫ g₀ and the same θ variation should matter much less. Two separations, same θ points as the single-ring sweep. Turns the closing claim into a result and ties the two halves of the paper together.

\section*{Discussion}

The results above show that a magnon hosted on one ring can be read out through a second ring with no physical connection to it, and that as the rings are separated, the magnon's interaction with the signal crossing the gap strengthens rather than weakens.
Let us now consider what this means for the wireless link as a whole.

\begin{figure}[htb]
\centering
\includegraphics[width=0.5\linewidth]{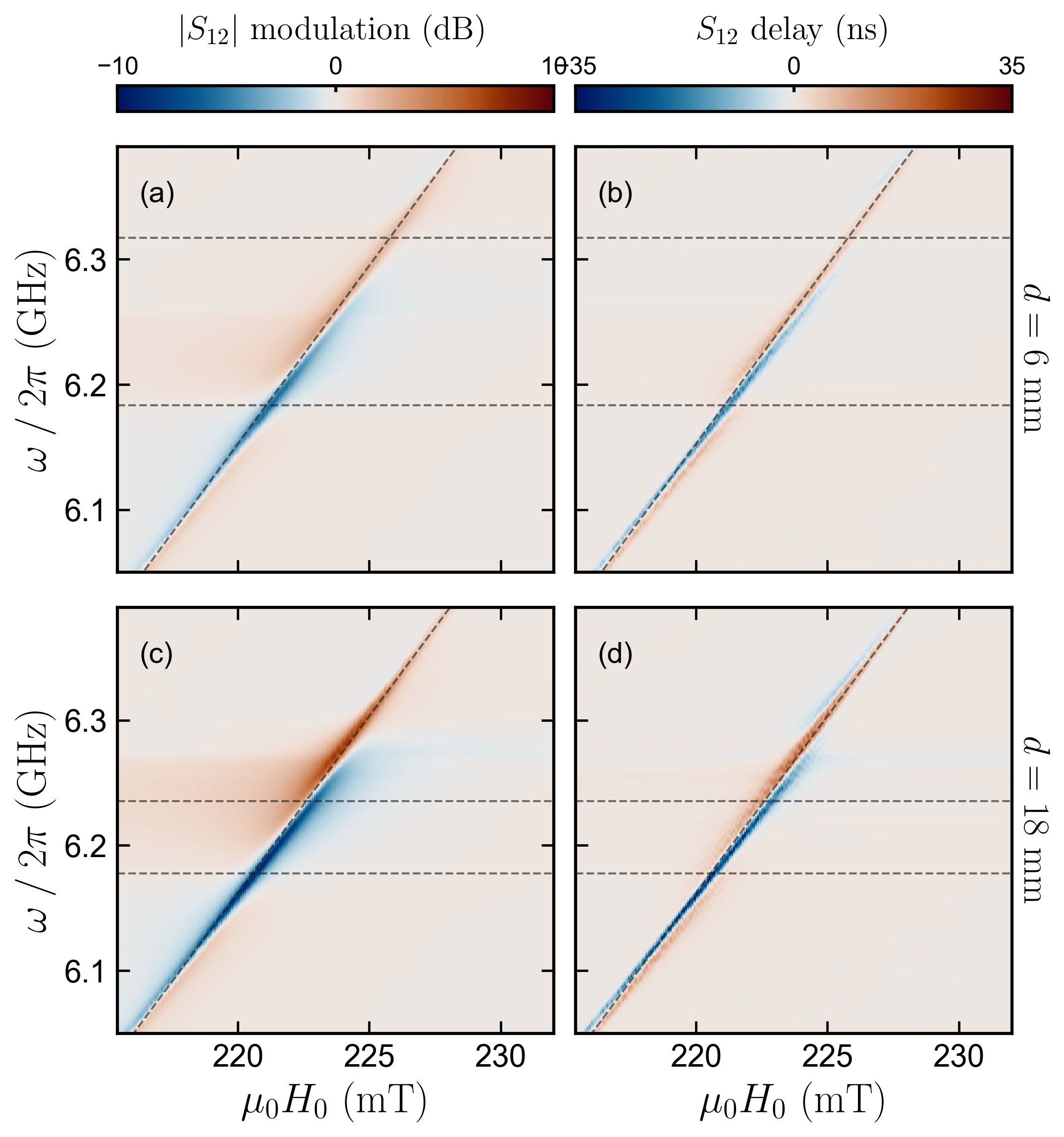}
\caption{\textbf{Magnon control of the wireless link.} Magnon-induced change in the transmission between the rings, referenced to the magnon-free response, at $d = 6$~mm (a),(b) and $d = 18$~mm (c),(d). (a),(c) Amplitude modulation of $|S_{12}|$. (b),(d) Group delay $\tau = -\mathrm{d}\phi/\mathrm{d}\omega$ contributed by the magnon, obtained after subtracting the off-resonant background. Diagonal dashed lines mark the bare magnon frequency, $\omega_0 = \gamma\mu_0 H_0$; horizontal dashed lines mark the hybridised ring modes.}
\label{fig:fig4}
\end{figure}

The link itself does not depend on the magnon. With the magnon far detuned, a transmission band persists out to $d = 18$~mm, the largest separation measured, so the magnon acts on a wireless link that already exists rather than creating one. What it does instead is redistribute the transmitted amplitude and phase around the avoided crossing. Referenced to the magnon-free level [Fig.~\ref{fig:fig4}(a),(c)], at $d = 6$~mm the magnon enhances the inter-ring transmission by up to $+3.1$~dB and suppresses it by up to $-5.2$~dB, rising to $+6.4$ and $-9.6$~dB at 18~mm. The modulation grows with separation for the same reason the magnon's coupling to the lower mode does: as the rings are drawn apart, the signal crossing the link resides increasingly in ring~2, where the magnon sits, so a larger share of it is shaped by the magnon. The two rings therefore form a wireless link that remains substantial out to 18~mm, across which a magnon hosted on only one of them controls the transmission by up to 16~dB peak-to-peak, tuned in situ by the static field.

This modulation is a steady-state effect, but the same system has a natural dynamical reading. The magnon linewidth is far narrower than that of the rings, so the magnon stores energy for $1/2\kappa_m \approx 20$~ns, against $\approx 1.3$~ns for the rings. Energy entering the link near the magnon resonance is therefore held in the magnetic system long after the ring fields have decayed. 
This appears directly in the phase of the transmission. The time taken to cross the link is given by the slope of the transmitted phase with frequency, $\tau = -\mathrm{d}\phi/\mathrm{d}\omega$; subtracting the off-resonant background leaves the delay contributed by the magnon alone [Fig.~\ref{fig:fig4}(b),(d)]. At every separation this delay changes sign across the magnon line. On one side, the signal is delayed by up to $17.8$~ns at $d = 18$~mm, consistent with energy being held in the magnon before being passed on to ring~1. On the other, the delay becomes negative, reaching $-31.9$~ns. % A negative delay does not mean that energy crosses the link faster: near an absorption feature, the leading edge of a pulse is transmitted preferentially over its trailing edge, so the peak of the transmitted envelope is advanced, a well-known consequence of resonant absorption. 
Both values % remain within the bound $1/\kappa_m \approx 41$~ns set by the magnon linewidth, and both 
are some three orders of magnitude longer than the free-space propagation time across the gap, $0.06$~ns. The delay is therefore set by resonant storage in the magnon rather than by the path length, and is tunable in situ with the static field. 
%In the time domain, this would appear as a long-lived tail in the energy arriving at ring~1, an effect accessible with pulsed excitation. 
%A magnonically controlled link of this kind is therefore tunable not only in how much energy crosses the gap, but in how long it takes to do so. 

At present, the exchange of energy between the magnon and the hybridised ring modes is limited by dissipation: a photon leaves the rings within $\approx 1.3$~ns, faster than the magnon can exchange energy with either hybridised mode. This is, however, a limitation of the present resonators rather than of the scheme. Reducing the ring linewidth below the magnon's coupling to each hybridised mode, would place each hybridised mode in the regime where the coupling directly sets the rate at which energy, and thus information, is exchanged between the magnon and the link.

In this regime, the redistribution of coupling observed in both reflection and transmission acquires a direct dynamical meaning. Since the magnon couples only to ring~2, its coupling to each hybridised mode, and hence the rate at which it exchanges energy with that mode, is set by how much of the mode resides on ring~2. The link does not create or destroy coupling, but shares it between two photonic modes that are each delocalised across the gap. The magnon remains coupled to both, with the separation setting how strongly, and the static field setting which of the two it is brought into resonance with. Combined with the long storage time of the magnon \cite{Serha_ultralong_2026}, this suggests that energy could be transferred into the magnon through the link, held there well beyond the lifetime of the rings, and later transferred out to ring~1, with the rate of each transfer set by the separation and the static field. Furthermore, extending our system into a two-arm geometry, in which the second feed addresses the standing-wave mode dark to the first allows for an additional degree of freedom. Transmission between the arms is therefore forbidden at lowest order and becomes finite only through backscattering or through the magnon. Evidence that the magnon remains coupled to the dark mode and that $\beta$ is finite is provided in Supplemental Note D.

The framework explored here offers potential for contactless transduction and scaling to resonator arrays. Though planar resonator systems are potentially limited in their use at room temperature due to low $Q_i$, their design and electromagnetic behaviour translate well to low temperature superconducting systems, including ring resonators \cite{sun2025superconducting}. For hybrid quantum systems, there is great interest in the integration of magnetic elements with superconducting circuits, and planar resonator designs are distinct from 3D geometries in their ability to integrate these elements, including, for example, metallic thin films which enable system control through the spin hall effect and can be directly metallised onto the structure \cite{hou2024electrical}. Vice versa, fabricating planar resonators onto insulating magnetic films can lead to strong coupling between photons and more complex spin textures and waves \cite{li2026strong}.

This architecture also has promise to be employed beyond a magnon read out mechanism.
For example, similar multimode resonator systems have recently demonstrated dynamically tuneable antimode exceptional points when under multi-drive control\cite{smith_exceptional_2026}. Our system also relates to a recent demonstration of a room-temperature source of correlated microwave magnon polariton pairs, based on a YIG film coupled to a printed-circuit-board microstrip resonator~\cite{wang_room-temperature_2026}. These are applications at room temperature that may not require the $Q_i$ factors demanded in potential quantum applications. 

In summary, we have demonstrated remote excitation and readout of cavity magnon-polaritons through a second planar ring resonator with no magnetic element and no wired connection to the one hosting the magnon, bringing the classical mechanism of near-field wireless energy transfer into cavity magnonics. The position of the YIG sphere on its ring sets which standing-wave pattern the magnon couples to, and the separation between the rings sets how that coupling is shared across the gap. As the rings are drawn apart, the magnon's coupling is not lost but concentrated into the mode residing on its own ring, so that its control over the signal crossing the link strengthens even as the link itself weakens. The result is a wireless link that remains substantial out to 18~mm, readable in both reflection and transmission, across which a single magnon mode modulates the transmission by up to 16~dB and delays it by tens of nanoseconds, both tuned in situ by the static field. Built from simple printed-circuit-board resonators, such near-field magnonic links turn the integration advantages of planar geometries into a new capability, offering a route towards contactless magnonic devices and arrays of resonators in which magnons are addressed and coupled without wired connections. 

% Limitations

$ $
\section*{Methods}

The experimental system %[Fig.~\ref{fig:remoteCouplingSetUp}(a)] 
consists of a planar microstrip ring of mean radius $r=4.5$~mm patterned on a Rogers RO4350B substrate with relative permittivity $\varepsilon_r=3.48$, substrate thickness $h=0.76$~mm, and 1~oz ($35~\mu$m) etched copper metallisation, fed through capacitive stubs. The ring has a width $W_{\rm ring}=1$~mm, while the coupling stubs have a length $H_{\rm stub}=0.6$~mm and are separated from the ring by a gap $g=0.2$~mm. The feedline has a width of 1.72~mm. 
For $r=4.5$~mm, the mean circumference is $2\pi r\simeq28.3$~mm. For the $1$-mm-wide microstrip, using $\varepsilon_r=3.48$, $h=0.762$~mm, and $w=1$~mm gives $\varepsilon_\text{eff}\simeq2.63$ \cite{Hammerstad1980, pozar_microwave_2011}. At $f=6.2$~GHz, this corresponds to a guided wavelength $\lambda_g\simeq29.8$~mm. Thus, $2\pi r/\lambda_g\simeq0.95$, indicating that approximately one guided wavelength fits around the ring. We therefore identify the observed resonance near $\omega_c/2\pi=6.2$~GHz as the $n=1$ resonance. 

The magnetic sample is a yttrium iron garnet (YIG) sphere of diameter 0.25~mm (Ferrisphere Inc.). It is mounted on a post secured to the PCB, positioning the sphere directly above the ring track, where it couples to the microwave magnetic field of the ring resonator. Measurements were repeated with the sphere relocated to different positions around the track.

\subsection*{Experimental Setup Details}
$ $

The scattering matrix $S_{ij}(\omega)$ ($i,j\in\{1,2\}$) was acquired acquired with a Rohde \& Schwarz ZNLE18 vector network analyser (VNA) at an output power of 0~dBm and an IF bandwidth of 500~Hz. A static bias field $\mathbf{H}_0$, oriented perpendicular to $\mathbf{h}_c$, was applied using a water-cooled GMW~3470 dipole electromagnet driven by a CAEN FAST-PS-M DC power supply. The VNA frequency sweep and the magnet current were controlled synchronously using \texttt{qkit}, an open-source measurement suite developed at the University of Glasgow~\cite{qkit}.

The electromagnet is current-driven, so the field axis was calibrated from the spectra themselves. At each current setting, the magnon resonance frequency was extracted from the spectra, away from the avoided crossing. It was then converted to field using the Kittel relation for a sphere, $\omega_m=\gamma\mu_0H_0$. A linear fit across the current sweep provides the current-to-field calibration. All measurements were performed at room temperature.

\subsection*{Simulation}

Finite-element simulations were performed in COMSOL Multiphysics (RF Module, Electromagnetic Waves, Frequency Domain interface) using the device geometry described above. The YIG sphere is modelled through its permeability tensor which, for a bias field along $\hat{z}$, takes the form
\begin{equation}
\bar{\mu}=\mu_0\begin{pmatrix}\mu_r & -i\kappa_r & 0\\ i\kappa_r & \mu_r & 0\\ 0 & 0 & 1\end{pmatrix},\qquad
\mu_r=1+\frac{\omega_0\omega_m}{\omega_0^2-\omega^2},\qquad
\kappa_r=\frac{\omega\,\omega_m}{\omega_0^2-\omega^2},
\end{equation}
where $\omega_m=\gamma\mu_0M_s$ and $\omega_0=\gamma\mu_0(H_0-N_{zz}M_s)+i\alpha\omega$. Here $\alpha$ is the Gilbert damping, and the demagnetising factor $N_{zz}=1/3$ for a sphere accounts for the reduction of the internal static field. The YIG parameters used are $\mu_0M_s=0.1758$~T, $\gamma/2\pi=28.0$~GHz/T, $\alpha=10^{-4}$, $\varepsilon_r=12.9$ and $\sigma=0$~\cite{stancil2009spi}. %

Data analysis and theoretical modelling were performed in Python 3.9.6, using NumPy (1.26.2) and SciPy (1.11.4) for curve fitting and signal processing, and Matplotlib (3.8.2). 

% circle-fit, input--output fit in Supplementary?

\section*{Data availability}

All experimental data presented in this study is available at DOI TBC %\href{http://dx.doi.org/10.5525/gla.researchdata.1792}{http://dx.doi.org/10.5525/gla.researchdata.1792}.

\section*{Acknowledgements}
This work was supported by the Science and Technology Facilities Council (STFC) under grant number UKRI600, the University of Manitoba, the Natural Sciences and Engineering Research Council of Canada (NSERC RGPIN 05011-18, RGPIN-2018-05012, and ALLRP 597808-24), and the Canadian Foundation for Innovation (CFI) John R. Evans Leaders Fund. 
M. A. Smith acknowledges funding from the Engineering and Physical Sciences Research Council, United Kingdom (EPSRC) Centre for Doctoral Training in Photonic Integration and Advance Data Storage (PIADS), under grant numbers EP/S023321/1.
A. Eden’s work was supported by the EPSRC Centre for Doctoral Training in Diversity-Led, Mission-Driven Research (DiveIn CDT), Grant Number EP/Y034902/1.
The work of M. Last, S. Howard, and A. Oravcová was conducted under the University of Glasgow 5$^{\text{th}}$ year Engineering course \emph{`Design Special Topic 5'}.
The authors also thank Andrew Phillips at the University of Glasgow for assistance with PCB fabrication.

\noindent For the purpose of open access, the authors have applied a Creative Commons Attribution
(CC BY) license to any Author Accepted Manuscript version arising from this submission.

\section*{Author Contributions}

AE, ML, SH, and AO performed initial resonator design, modeling and experimental characterisation, and with input from MAS, AJ, and RM.
AJ and AE optimised the experimental set-up with theoretical input from RDM and performed the measurements with YIG with support from MAS and input from RLS and RM. 
AJ carried out data analysis with contributions from AE and MAS. 
RDM derived the second quantised model. 
AJ wrote the manuscript with input from and discussions with all coauthors. 
MAS conceived the experiment and MAS, RLS, and RM supervised the project.

\section*{Competing Interests}

The authors declare no competing interests.

\bibliography{sample}
\newpage
\clearpage
\renewcommand{\theequation}{S\arabic{equation}}
\renewcommand{\thetable}{S\arabic{table}}
\renewcommand{\thefigure}{S\arabic{figure}}
\setcounter{equation}{0}
\setcounter{table}{0}
\setcounter{figure}{0}
\setcounter{page}{0}
\maketitle

\noindent{\huge\textbf{Supplementary Information for:\vspace{0.3cm}\\Wireless readout and control of cavity \vspace{0.3cm}\\ 
magnon-polaritons via photonic modes \vspace{0.3cm}\\  delocalised across a gap}}
\vspace{0.5cm}

\noindent \textbf{\large Alban Joseph\textsuperscript{1,*}, Ryan D McKenzie\textsuperscript{2}, Amy Eden\textsuperscript{1}, Morven Last\textsuperscript{1}, Shannon Howard\textsuperscript{1}, Alžbeta Oravcová\textsuperscript{1}, Mawgan A Smith\textsuperscript{1}, Robert L Stamps\textsuperscript{2}, and Rair Mac\^edo\textsuperscript{1, $\dag$}}
\vspace{0.5cm}

\noindent \textsuperscript{1}James Watt School of Engineering, Electronics \& Nanoscale Engineering Division, University of Glasgow, Glasgow
G12 8QQ, United Kingdom

\noindent \textsuperscript{2}Department of Physics and Astronomy, University of Manitoba, 30A Sifton Road, Winnipeg R3T 2N2, Manitoba, Canada

\noindent $^*$Alban.Joseph@glasgow.ac.uk \\ 
\noindent $^\dag$Rair.Macedo@glasgow.ac.uk
\thispagestyle{empty}
\newpage

\section*{Supplementary Note A: Input--Output Theory of the Empty Resonator}
$ $
\label{app:io}

\begin{figure}[htp]
	\centering       
	\mbox{\includegraphics[width=8.5cm]{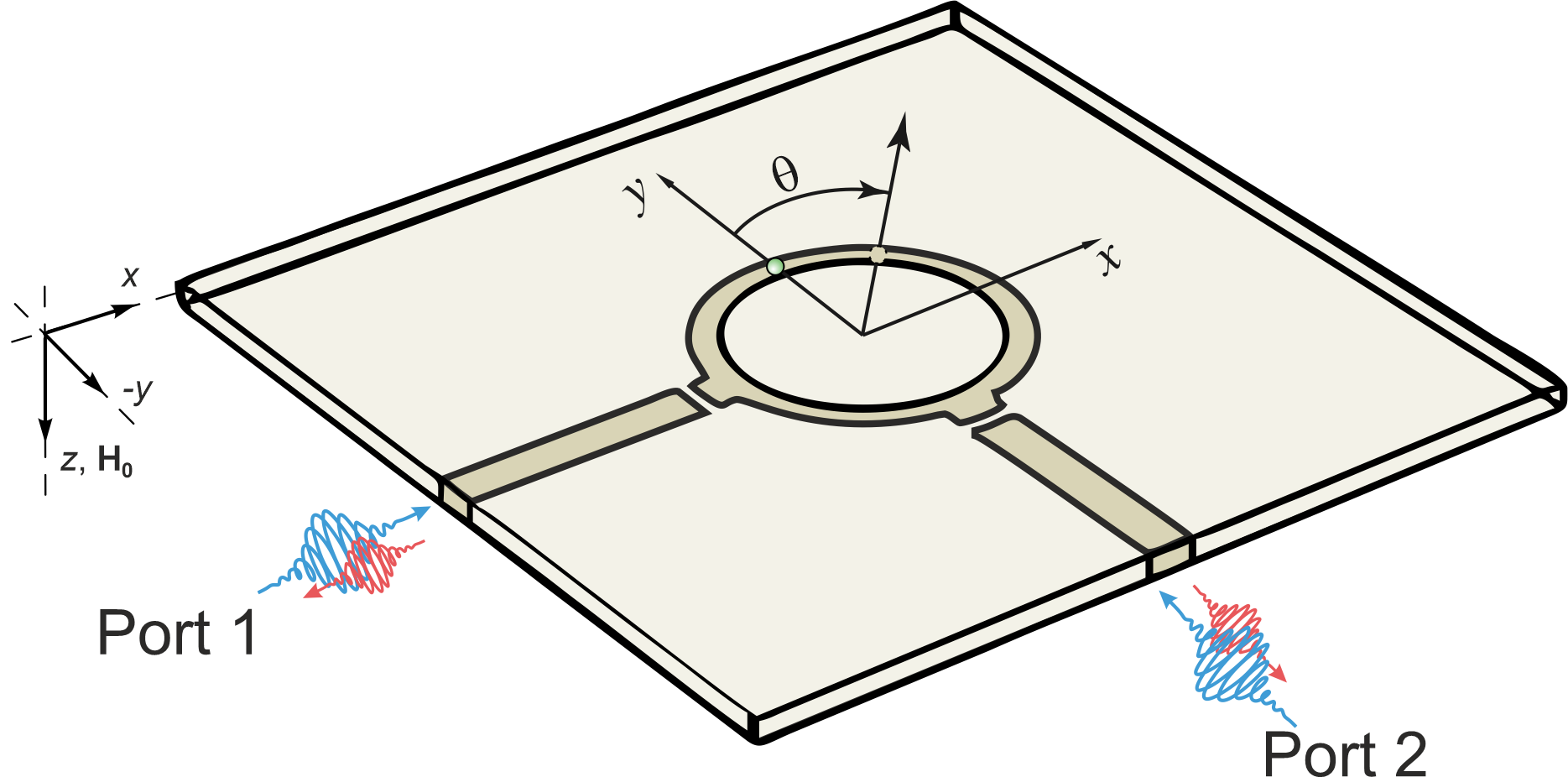}}
	\hspace{5px}
	\mbox{\includegraphics[width=8.5cm]{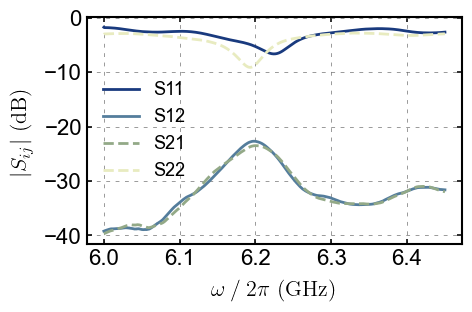}}
	\caption{\textbf{full resonator response characterisation.}(a) Two-arm resonator geometry allowing for backscatering between CW and CCW modes to be separately characterised and (b) full scattering matrix parameters measured using the set up in (a) where S11 and S22 are the reflection coefficients measured/excited at Port 1 and Port 2 respectively and S21(S12) is the transmission coefficient measured at Port 2(1) and excited at Port 1(2).}
	\label{fig:diagram}
\end{figure}

In this section, we use input-output theory \cite{GardinerCollett, GardinerZollerBook} to determine the response of the empty resonator. To begin, we consider a coherent $\beta = |\beta| e^{i \varphi_{\beta}}$, and dissipative $\beta_d$, coupling between the CW ($+$) and CCW ($-$) modes present in the ring resonator ($\hbar=1$)
\begin{align}
	\mathcal{H}_{sys} = \widetilde{\omega}_a a_{+}^{\dag} a_{+} 
    + \widetilde{\omega}_a a_{-}^{\dag} a_{-} +
	\beta a_{+}^{\dag}a_{-}+ \beta^* a_{-}^{\dag}a_{+} +
	i\beta_d(a_{+}^{\dag}a_{-}+a_{-}^{\dag}a_{+}),
\end{align}
with $\widetilde{\omega}_a = \omega_a - i \kappa_{\rm int}$, where $\omega_a$ is the resonator frequency, and $\kappa_{\rm int}$ is the half-width-half-maximum of the mode due to its internal damping. The coherent and dissipative couplings are the result of backscattering from the resonator stubs. The $B$-fields generated by the CW and CCW modes are 
\begin{align}
	B_{cw}(r,z,\theta) = B(r,z) e^{i(\theta+\pi/2)} a_{+}  + H.c.
	\qquad\qquad
	B_{ccw}(r,z,\theta) = B(r,z) e^{-i(\theta+\pi/2)} a_{-}  + H.c.
\end{align}
The factor of $e^{\pm i \theta}$ accounts for the angular dependence of the field around the ring for the $n=1$ mode with $\theta=0$ corresponding to the 12 O'clock position, and $B(r,z)$ accounts for the strength of the field at the radial position $r$ and height $z$. As we restrict our attention to a YIG sphere placed directly on the ring, with fixed $r$ and $z=0$, we assume $B(r,z)=B$ in what follows. 

Rather than working with the CW and CCW modes, it is convenient to work in a standing-wave basis
\begin{align}
	a_s = \frac{1}{\sqrt{2}}(a_{+}+a_{-}) \qquad\qquad 
	a_c = \frac{1}{\sqrt{2}}(a_{+}-a_{-}).
\end{align}
In this basis one finds
\begin{align}
	\mathcal{H}_{sys} = \widetilde{\omega}_s a_s^{\dag} a_s 
    + \widetilde{\omega}_c a_c^{\dag} a_c 
	+ g_{cs} (a_s^{\dag} a_c - a_c^{\dag} a_s)
\end{align}
where
\begin{align}
\label{eq:params}
	\widetilde{\omega}_s = \widetilde{\omega}_a+\text{Re}[\beta]+i\beta_d \qquad 
	\widetilde{\omega}_c = \widetilde{\omega}_a-\text{Re}[\beta]-i\beta_d \qquad\qquad 
    g_{cs} = -i\text{Im}[\beta].
\end{align}
Each part of the backscattering plays a distinct role: $\mathrm{Re}[\beta]$ splits the two modes, $\mathrm{Im}[\beta]$ mixes them, and $\beta_d$ modifies their linewidths. The $B$-field operator in the standing wave basis is
\begin{align}
\label{eq:qB}
	B(\theta) = B_{cw}(\theta) + B_{ccw}(\theta) =  
	\sqrt{2} B [i (a_c-a_c^{\dag}) \cos \theta - (a_s + a_s^{\dag}) \sin \theta] .
\end{align}
Having determined the system Hamiltonian and resulting $B$-field in the standing wave basis, we now consider the ring of the main text with the port~1 feed arm located at $\theta=-\pi/2$, and a second feed arm connected to port~2 a quarter of a wavelength around the ring from the first (i.e. $\theta=\pi$ for $n=1$). As discussed in the main text, each capacitive stub drives the standing-wave pattern with a magnetic-field node at its own position. Port~1 therefore addresses the $\cos(n\theta)$ pattern and port~2 the $\sin(n\theta)$ pattern, and each pattern is dark to the opposite port. In the absence of any coupling between the two patterns, the ports are thus isolated and $S_{21} = S_{12} = 0$.

The bath Hamiltonian is
\begin{align}
	\mathcal{H}_{bath} = \sum_{n=1,2} \int_{-\infty}^{\infty}  
	\omega b_n^{\dag}(\omega) b_n(\omega)\ d\omega,
\end{align}
where $b_1^{\dag}$ is associated with the left stub at $\theta=-\pi/2$, and $b_2^{\dag}$ is associated with the bottom stub at $\theta=\pi$. For a single stub located at $\theta=\theta_0$ the system-bath interaction in the CW/CCW basis is
\begin{align}
	\mathcal{H}_{int,0} = i \int_{-\infty}^{\infty} d\omega\  
	[\kappa_0 b_0^{\dag}(\omega) a_{+} - \kappa_0^* b_0(\omega) a_{+}^{\dag}] -
	i \int_{-\infty}^{\infty} d\omega\  
	[\kappa_0^* b_0^{\dag}(\omega) a_{-} - \kappa_0 b_0(\omega) a_{-}^{\dag}],
\end{align}
where $\kappa_0 = \kappa e^{i (\theta_0+\pi/2)}$. The minus sign between the CW and CCW terms accounts for a $180^{\circ}$ phase shift between these modes. In the standing-wave basis the interaction is
\begin{align}
	\mathcal{H}_{int,0}= i \int_{-\infty}^{\infty} d\omega\  
	[\kappa_s^* b_0^{\dag}(\omega) a_s - \kappa_s b_0(\omega) a_s^{\dag}] +
	i \int_{-\infty}^{\infty} d\omega\  
	[\kappa_c^* b_0^{\dag}(\omega) a_c - \kappa_c b_0(\omega) a_c^{\dag}]
\end{align}
where
\begin{align}
	\kappa_s^* = i\sqrt{2} \kappa \cos \theta_0
	\qquad\qquad \kappa_c^* = -\sqrt{2} \kappa \sin \theta_0.
\end{align}
With the resonator stubs positioned at $\theta=-\pi/2$ and $\pi$, the system-bath interaction in the standing wave basis is 
\begin{align}
	\mathcal{H}_{int} =
    i \int_{-\infty}^{\infty} d\omega \ \boldsymbol{b}^{\dag} K^{\dag} \boldsymbol{a} 
	- i \int_{-\infty}^{\infty} d\omega \  \boldsymbol{a}^{\dag} K \boldsymbol{b},
\end{align}
where $\boldsymbol{b}^{\dag} = \begin{bmatrix} b_1^{\dag} & b_2^{\dag}  \end{bmatrix}$ and $\boldsymbol{a}^{\dag} = \begin{bmatrix} a_c^{\dag} & a_s^{\dag}  \end{bmatrix}$, and the coupling matrix is given by
\begin{align}
	K^{\dag} =  \begin{bmatrix}
		\sqrt{2} \kappa & 0 \\ 0 & -i\sqrt{2} \kappa \\
	\end{bmatrix} \qquad\qquad
	\pi K K^{\dag} = \begin{bmatrix}
		2\pi \kappa^2 & 0 \\ 0  & 2\pi \kappa^2 \\
	\end{bmatrix} \equiv \begin{bmatrix}
		\kappa_{\rm ext} & 0 \\ 0  & \kappa_{\rm ext} \\
	\end{bmatrix} .
\end{align}
For simplicity, we assume the coupling constant $\kappa$ is the same at both resonator ports.

Following standard input-output theory, the equations of motion for the cavity modes are
\begin{align}
	\dot{a}_c &= -i z_c a_c - ig_{cs}^* a_s - \sqrt{2 \kappa_{\rm ext}} \ b_{1,in} 
	\qquad\qquad\qquad
	\dot{a}_s = -i z_s a_s  - ig_{cs} a_c - i \sqrt{2 \kappa_{\rm ext}} \ b_{2,in},
\end{align}
where $z_{c(s)}=\widetilde{\omega}_{c(s)}-i\kappa_{\rm ext}$ includes damping from the resonator ports, and the boundary conditions are
\begin{align}
	\boldsymbol{b}_{out}-\boldsymbol{b}_{in} = \sqrt{2\pi} K^{\dag} \boldsymbol{a}. 
\end{align}
In frequency space one finds
\begin{align}
	i \Omega  \boldsymbol{a} = \sqrt{2\pi} K \boldsymbol{b}_{in} \qquad\text{with}\qquad
	\Omega = \left[ \begin{array}{cc}
		\omega-z_c & -g_{cs}^* \\ -g_{cs} & \omega-z_s  \\
	\end{array} \right].
\end{align}
Defining $D_c^{-1}=\omega-z_c$ and $D_s^{-1}=\omega-z_s$, and inverting the time evolution matrix, one finds $\boldsymbol{a} = \chi K \boldsymbol{b}_{in}$, with
\begin{align}
	\chi=-i \sqrt{2\pi} \Omega^{-1} 
    = \frac{-i \sqrt{2\pi}}{\text{det}[\Omega]} 
	\begin{bmatrix}
		D_s^{-1} & g_{cs}^* \\ g_{cs} & D_c^{-1}  \\
	\end{bmatrix} = \begin{bmatrix}
	\chi_c & \chi_{cs} \\ 
    \chi_{sc} & \chi_s  \\
	\end{bmatrix} 
\end{align}
The $S$-matrix, $\boldsymbol{b}_{out} = S \boldsymbol{b}_{in}$, follows from the boundary conditions
\begin{align}
	\boldsymbol{b}_{out}=\boldsymbol{b}_{in}+\sqrt{2\pi} K^{\dag} \boldsymbol{a}
	=\boldsymbol{b}_{in} + \sqrt{2\pi} K^{\dag} \widetilde{\chi} K \boldsymbol{b}_{in}
	\qquad\Rightarrow\qquad S=\mathbb{I}+ \sqrt{2\pi} K^{\dag} \widetilde{\chi} K
\end{align}
One finds
\begin{align}
	S_{12} = S_{21} =  \frac{i 2 \kappa_{\rm ext} 
    \text{Im}[\beta]}{(\omega-z_c)(\omega-z_s)-(\text{Im}[\beta])^2 },
\end{align}
so the transmission is reciprocal, as expected, and $S_{21}=S_{12}=0$ if there is no mixing of the standing waves (i.e., if $\text{Im}[\beta] = 0$). The reflection coefficients are given by
\begin{align}
	S_{11} = \frac{(\omega-z_c) (\omega-z_s - i2\kappa_{\rm ext})-(\text{Im}[\beta])^2}
	{(\omega-z_c) (\omega-z_s)-(\text{Im}[\beta])^2} 
	\\ \nonumber
	S_{22} = \frac{(\omega-z_c- i2\kappa_{\rm ext}) (\omega-z_s)-(\text{Im}[\beta])^2}
	{(\omega-z_c) (\omega-z_s)-(\text{Im}[\beta])^2}
\end{align}
The poles of the $S$-matrix elements are the modes of the system. With $\text{Im}[\beta] \neq 0$, the standing waves are mixed and the modes are split; however, in the experiment, $\text{Im}[\beta]$ is small and the splitting of the modes is not apparent in the transmission spectra. Rather, the measured transmission appears broadened. One may, however, use the reflection spectra to determine both the coherent and dissipative couplings between the modes as shown below. 

\subsection*{Extraction of the backscattering}

The three components of the backscattering leave distinct signatures in the measured spectra (Fig. \ref{fig:ERthetaJd}). The offset between the minima of $S_{11}$ and $S_{22}$, approximately $35.5$~MHz, sets $2|\beta|$, giving $|\beta|/2\pi \approx 17.75$~MHz. The peak transmission, $|S_{21}| \approx -22$~dB, determines the phase: a real $\beta$ ($\varphi_\beta = 0, \pi$) gives no transmission, while $\varphi_\beta \approx 180^\circ \pm 17^\circ$, with $\kappa_{\rm ext}/(2\pi)=10$ MHz and $\kappa_{\rm int}/(2\pi)=25$ MHz, reproduces the measured level. The two values of $\varphi_\beta$ are indistinguishable in the empty-ring spectra, but with the magnon included one finds that $\varphi_\beta = 197^\circ$ reproduces the measured antimode spectrum (see Supplementary Note B). Finally, the [3--4]~dB difference in depth between the $S_{11}$ and $S_{22}$ minima, together with the asymmetry in their linewidths, gives $\beta_d/2\pi \approx 5$~MHz. 

\begin{figure}[htp]
	\centering       
	\mbox{\includegraphics[width=8.5cm]{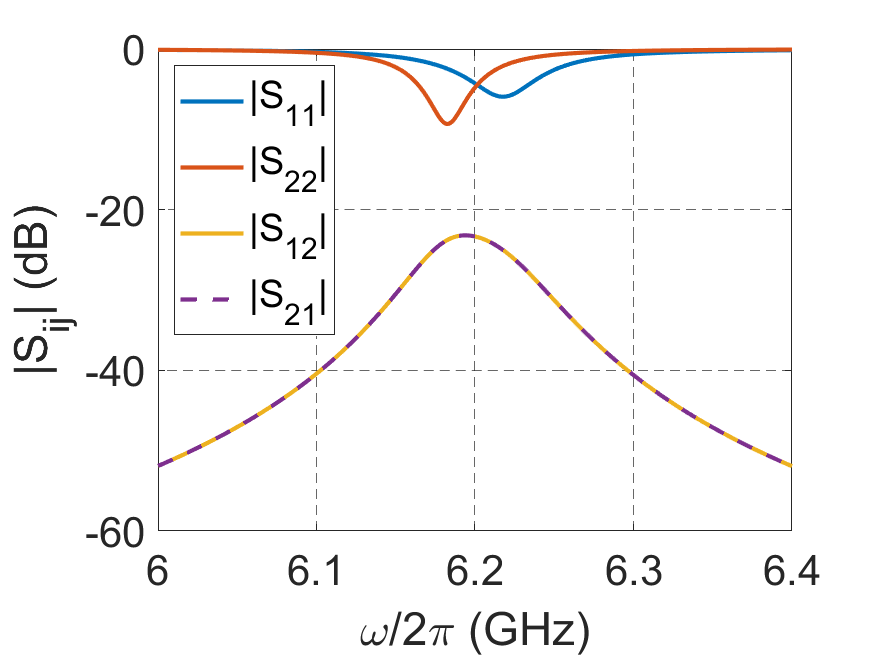}}
	\hspace{5px}
	\mbox{\includegraphics[width=8.5cm]{figures/supplementary/S1b.png}}
	\caption{Calculated and measured transmission spectra. On the left we assume the standing wave modes have frequency $\omega_c=6.21$ GHz, with intrinsic dampings (HWHM) of $\kappa_{\rm int} = 25$ MHz and external dampings of $\kappa_{\rm ext} = 10$ MHz. The phase of the coherent part of the coupling is set to $\theta=197^{\circ}$ and its magnitude is $|J|/(2\pi)=17.75$ MHz. The strength of the dissipative component of the coupling is $J_d=5$ MHz. This provides good agreement with the measured transmission spectra shown on the right.}
	\label{fig:ERthetaJd}
\end{figure}

% \begin{figure}[htp]
% 	\centering       
% 	{\includegraphics[width=8.5cm]{figures/supplementary/ERrandtC_ERExpt.png}}
% 	\caption{Calculated (bottom) and measured (top) transmission spectra. On the bottom we assume the standing wave modes have frequency $\omega_c=6.21$ GHz, with intrinsic dampings (HWHM) of $\kappa_{\rm int} = 25$ MHz and external dampings of $\kappa_{\rm ext} = 10$ MHz. The phase of the coherent part of the coupling is set to $\theta=197^{\circ}$ and its magnitude is $|J|/(2\pi)=17.75$ MHz. The strength of the dissipative component of the coupling is $J_d=5$ MHz. This provides good agreement with the measured transmission spectra shown on the right.}
% 	\label{fig:ERthetaJd}
% \end{figure}

\cleardoublepage

%%%%%%%%%%%%%%%%%%%%%%%%%%%%%%%%%%%%
\section*{Supplementary Note B: Magnon-Photon Coupling and Input-Output Theory of the Loaded Resonator}
\label{app:mp}
%%%%%%%%%%%%%%%%%%%%%%%%%%%%%%%%%%%%%

Having characterized the empty resonator, we now include a YIG sphere in the model. The Hamiltonian of a magnetic moment polarized in the $x$-direction, subject to an ac field in the $y$-direction, is
\begin{align}
	\mathcal{H}_{mp} = -\mu_x B_y \cos(\omega t)
\end{align}
For the case at hand, the quantized expression for the ac field is [Eq. (\ref{eq:qB})]
\begin{align}
	B_y \cos(\omega t) \rightarrow \langle B(\theta) \rangle = 
	\sqrt{2} B [i (\langle a_c \rangle - \langle a_c^{\dag} \rangle) \cos \theta 
    - ( \langle a_s \rangle + \langle a_s^{\dag} \rangle) \sin \theta],
\end{align}
where the time dependence is implicit in the photon operators. To leading order in the magnon operators, the quantized magnetic moment is given by
\begin{align}
\mu_x = \gamma \langle S^x \rangle \propto m+m^{\dag}.
\end{align}
This leads to the quantized magnon-photon Hamiltonian
\begin{align}
	\mathcal{H}_{mp} = 
	-g_c (a_c-a_c^{\dag}) (m+m^{\dag}) + g_s (a_s+a_s^{\dag}) (m+m^{\dag}).
\end{align}
where
\begin{align}
\label{eq:gcgs}
g_c=-ig_0 \cos \theta \qquad\text{and}\qquad g_s=-g_0 \sin \theta.
\end{align}
Dropping the counter-rotating terms, the system Hamiltonian with the magnon included is
\begin{align}
	\mathcal{H}_{sys} = \tilde{\omega}_{s} a_s^{\dag} a_s 
    + \tilde{\omega}_{c} a_c^{\dag} a_c + \tilde{\omega}_0 m^{\dag}m
	+ [g_{cs} a_s^{\dag} a_c + g_c a_c^{\dag} m + g_s a_s^{\dag} m + H.c.],          
\end{align}
where $\tilde{\omega}_0=\omega_0-i\kappa_m$ is the complex magnon mode frequency, and $\tilde{\omega}_{s}$, $\tilde{\omega}_{c}$, and $g_{cs}$ are given in Eq. (\ref{eq:params}). The bath Hamiltonian and the system-bath interaction are the same as in Supplementary Note A.

Following standard input-output theory, the equations of motion for the system operators are
\begin{align}
	\dot{a}_c &= -i z_c a_c -i g_{cs}^* a_s-i g_c m - \sqrt{2\kappa_{\rm ext}} \ b_{1,in} 
	\\ \nonumber
	\dot{a}_s &= -i z_s a_s -i g_{cs} a_c -i g_s m  - i \sqrt{2\kappa_{\rm ext}} \ b_{2,in}  
	\\ \nonumber
	\dot{m} &=  -i z_m m - i g_s^* a_s - i g_c^* a_c,
\end{align}
where the complex mode frequencies, $z_c$ and $z_s$, now include damping from the resonator ports and $z_m=\widetilde{\omega}_0$. Transforming to frequency space, the equation of motion for the magnon mode is
\begin{align}
	m = \chi_m (g_c^* a_c + g_s^* a_s) \qquad\text{where}\qquad 
	\chi_m^{-1} = \omega-\omega_0 + i\kappa_m.
\end{align}
After eliminating the magnon from the equations of motion, one finds the time evolution of the photonic modes is governed by
\begin{align}
	i \Omega  \boldsymbol{a} = \sqrt{2\pi} K \boldsymbol{b}_{in} 
    \qquad\text{with}\qquad
    \boldsymbol{a} = \begin{bmatrix} a_c \\ a_s \end{bmatrix} \quad
    \boldsymbol{b}_{in} = \begin{bmatrix} b_{1,in} \\ b_{2,in} \end{bmatrix}
\end{align}
and
\begin{align}
    K = \begin{bmatrix} \sqrt{2} \kappa & 0 \\ 0 & i \sqrt{2} \kappa \\ \end{bmatrix}
    \qquad\qquad
    \Omega = \begin{bmatrix}
		D_{cm}^{-1} & -\widetilde{g}_{cs} \\ -\widetilde{g}_{sc} & D_{sm}^{-1}  \\
	\end{bmatrix},
\end{align}
where the cavity mode response functions are
\begin{align}
    D_{cm(sm)}^{-1}(\omega) = \omega- z_{c(s)} - |g_{c(s)}|^2 \chi_m 
\end{align}
and the coupling parameters are
\begin{align}
	\widetilde{g}_{cs}(\omega) = g_c g_s^* \chi_m(\omega)  + g_{cs}^* 
    \qquad\text{and}\qquad
	\widetilde{g}_{sc}(\omega) = g_s g_c^* \chi_m(\omega)  + g_{cs}.
\end{align}
The modes of the system follow from the zeros of the determinant of the time evolution matrix $\text{det}[\Omega] = D_{cm}^{-1} D_{sm}^{-1} -\widetilde{g}_{cs}\widetilde{g}_{sc}=0$. Inverting the time evolution matrix $\Omega$, gives the cavity mode susceptibility $\boldsymbol{a} = \chi K \boldsymbol{b}_{in}$, where
\begin{align}
	\chi = -i \sqrt{2\pi} \Omega^{-1}
	= \begin{bmatrix}
	\chi_c & \chi_{cs} \\  \chi_{sc}  & \chi_s  \\
	\end{bmatrix}.
\end{align}
In terms of the cavity response functions and the couplings, the components of the susceptibility tensor are
\begin{align}
	\chi_{c(s)} = \frac{-i\sqrt{2\pi} D_{sm(cm)}^{-1}}{ D_{cm}^{-1} D_{sm}^{-1} - 
    \widetilde{g}_{cs}\widetilde{g}_{sc}} 
	\qquad\qquad
	\chi_{cs(sc)}  = \frac{-i\sqrt{2\pi} \widetilde{g}_{cs(sc)}}
	{D_{cm}^{-1}D_{sm}^{-1} - \widetilde{g}_{cs}\widetilde{g}_{sc}}. 
\end{align}
As for the empty resonator, one may use the boundary conditions and the cavity mode susceptibility to determine the $S$-matrix.

The $S$-matrix for this two port, three mode, system is given by
\begin{align}
	S = \mathbb{I} + \sqrt{2\pi} K^{\dag} \chi K
    =\begin{bmatrix}
		1+\sqrt{8\pi}\kappa^2 \chi_c & i\sqrt{8\pi}\kappa^2 \chi_{cs} \\ 
        -i \sqrt{8\pi} \kappa^2 \chi_{sc} & 1+\sqrt{8\pi}\kappa^2 \chi_s  \\
	\end{bmatrix}.
\end{align}
The transmission $S$-matrix elements are 
\begin{align}
    \label{eq:trans}
	S_{12} = S_{21} = i \kappa_{\rm ext} 
    \frac{  g^2 \sin (2 \theta) \chi_m  + 2\text{Im}[J]}
	{D_{cm}^{-1}D_{sm}^{-1} - \widetilde{g}_{cs}\widetilde{g}_{sc}},
\end{align}
so, counter-intuitively, the system is reciprocal regardless of the placement of the YIG sphere. The numerator of Eq. (\ref{eq:trans}) explicitly shows two distinct transmission pathways. The first term is a magnon-mediated conversion between the standing waves, which is maximal at $\theta = \pm \pi/4$ and vanishes at $\theta = 0, \ \pm \pi/2$ and $\pi$, where the sphere sits at a node of one of the standing waves. The second term is the conversion due to backscattering, which is independent of the sphere position. 

The reflection $S$-matrix elements are
\begin{align}
	S_{11} &= \frac{(D_{cm}^{-1}-i2\kappa_{\rm ext}) D_{sm}^{-1}-\widetilde{g}_{cs}\widetilde{g}_{sc}}
	{D_{cm}^{-1}D_{sm}^{-1}-\widetilde{g}_{cs}\widetilde{g}_{sc}}
	\qquad\qquad
	S_{22} = \frac{D_{cm}^{-1} (D_{sm}^{-1}-i2\kappa_{\rm ext})-\widetilde{g}_{cs}\widetilde{g}_{sc}}
	{D_{cm}^{-1}D_{sm}^{-1}-\widetilde{g}_{cs}\widetilde{g}_{sc}}.
\end{align}
Identifying features of the reflection spectra with the poles and zeros of $S_{11}$ and $S_{22}$ is not straightforward as there are multiple, nearly degenerate resonances and antiresonances. One can, however, use these expressions to model the experimental data and to extract the strength of the magnon-photon coupling from the splitting of the strong antiresonance observed in the data as shown in the Input-Output Theory of the Cavity Reflection below.

The theory presented above, with mixing of the sine and cosine cavity modes due to backscattering and the magnon, reproduces the experimental resonator data. To gain a better understanding of the effect of the magnon's position, we neglect backscattering in which case the magnon mediated couplings between the sine and cosine modes are
\begin{align}
	\widetilde{g}_{cs} = \frac{ig_0^2   \sin (2\theta) \chi_m}{2}   
    \qquad\text{and}\qquad
	\widetilde{g}_{sc}(\omega) = \frac{-ig_0^2   \sin (2\theta) \chi_m}{2},
\end{align}
and the reflection from port one is
\begin{align}
\label{eq:S11simp}
	S_{11} = 1-\frac{i2\kappa_{\rm ext}}{ D_{cm}^{-1}  - 
    \widetilde{g}_{cs}\widetilde{g}_{sc} D_{sm}}
    =1-\frac{i2\kappa_{\rm ext}}{ \omega- z_{c} - \frac{g_0^2 \cos^2 \theta}{\omega-z_m}   - 
    \frac{g_0^4   \sin^2 (2\theta) \chi_m^2/4}
    {\omega- z_s - \frac{g_0^2 \sin^2 \theta}{\omega-z_m}}}.
\end{align}
In the final term of this expression, we see that the driven resonator mode $z_c$ couples directly to the magnon with strength $g_0 |\cos \theta|$. There is also an indirect coupling between the driven cavity mode and the orthogonal cavity mode, $z_s$, mediated by the magnon. The strength of this coupling is largest at the resonant frequency of the magnon ($\chi_m$ is large), and when $\sin 2 \theta = 1$, i.e., where the magnon couples equally to the sine and cosine modes. The orthogonal cavity mode is then coupled back to the magnon with strength $g_0 |\sin \theta|$. In the absence of backscattering, with $\theta=0$ or $\pi$, the orthogonal cavity mode is decoupled from the rest of the system, and the reflection spectrum reduces to
\begin{align}
	S_{11} =1-\frac{i2\kappa_{\rm ext}}{ \omega- z_{c} - \frac{g_0^2}{\omega-z_m}}.
\end{align}
This simply describes reflection when a driven cavity mode is coupled directly to a magnon. At the field nodes, $\theta = \pi/2 \ \text{or} \ 3\pi/2$, the magnon is unable to couple to the driven cavity mode and $S_{11}$ reduces to the bare cavity response.

\begin{figure}[h]
  \centering
  \includegraphics[width=0.9\linewidth]{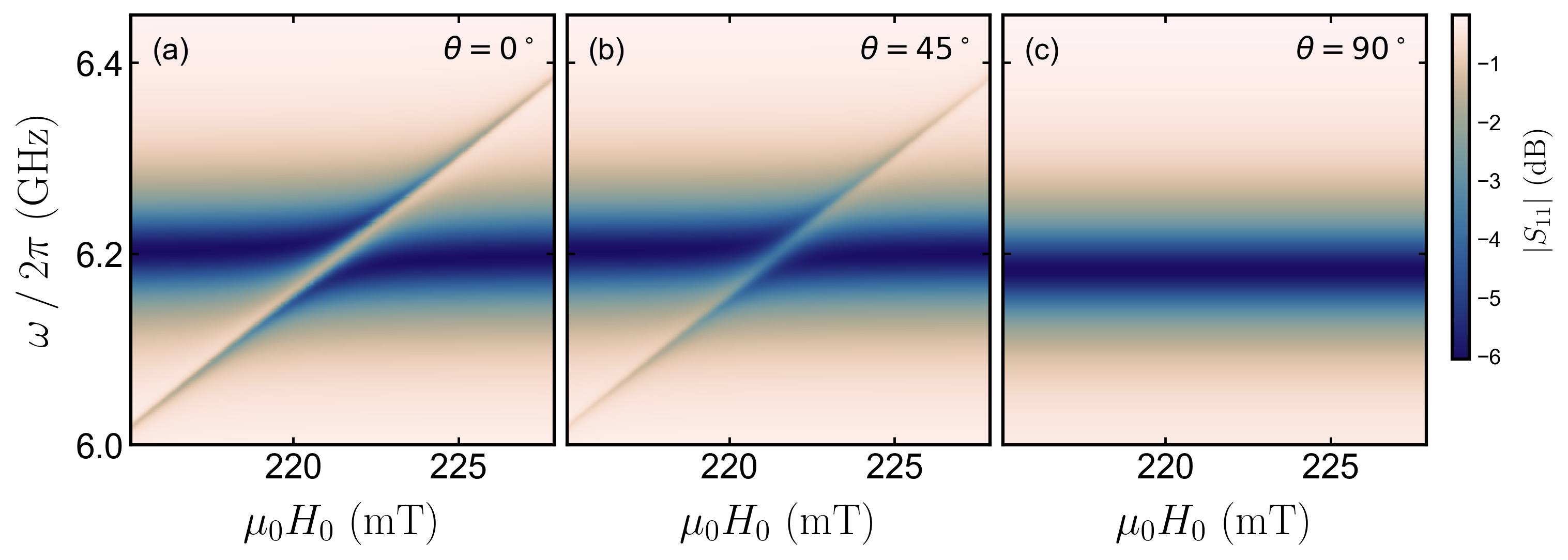}
  \caption{\textbf{Input--output model of the reflection spectrum as a function of the YIG position $\theta$.} $|S_{11}|$ computed from Eq.~\eqref{eq:S11} for (a) $\theta = 0^\circ$, (b) $\theta = 45^\circ$, and (c) $\theta = 90^\circ$, using the linewidths and coupling amplitude $g_0$ obtained independently from the loaded-resonator fit, with $g_c = g_0|\cos\theta|$ and $g_s = g_0|\sin\theta|$ coupling the magnon to the bright and dark cavity modes respectively.}
  \label{fig:trackPosition_heatmaps_model}
\end{figure}

This behaviour is demonstrated in Fig. \ref{fig:trackPosition_heatmaps_model}, which shows $S_{11}$ [Eq. (\ref{eq:S11simp}), or Eq. (\ref{eq:S11}) of the main text] with the YIG sphere positioned at $\theta = 0^\circ, \ 45^\circ$ and $90^\circ$. The parameters used are consistent with the experimental parameters extracted in the main text with the bare magnon-photon coupling being $g_0 \approx 10$ MHz. As $\theta$ rotates from $0^\circ$ to $90^\circ$ the avoided crossing collapses even though $g_0$ is unchanged because the magnon rotates off the bright mode, which radiates into the measurement port, and onto the dark mode, which does not --- reproducing the loss of visible splitting seen in the measured spectra. One may compare this to the spectra simulated in COMSOL shown in Fig.~\ref{fig:fig1} in which one sees evidence of the magnon even at $\theta=90^\circ$ due to backscattering. A more detailed comparison of the coupled mode model and the experimental data, including the effects of backscattering and dissipative couplings, is provided in Supplemental Note D.

\newpage
%%%%%%%%%%%%%%%%%%%%%%%%%%%%%%%%%%%%%%%%%%%%%%%%%%%%%%%%%%%%%%%%%%%%%%%%%%%%%%%%%%%%
\subsection*{Position Dependence of the Magnon--Photon Coupling}
%%%%%%%%%%%%%%%%%%%%%%%%%%%%%%%%%%%%%%%%%%%%%%%%%%%%%%%%%%%%%%%%%%%%%%%%%%%%%%%%%%%%

\begin{figure}[h]
  \centering
  \includegraphics[width=0.5\linewidth]{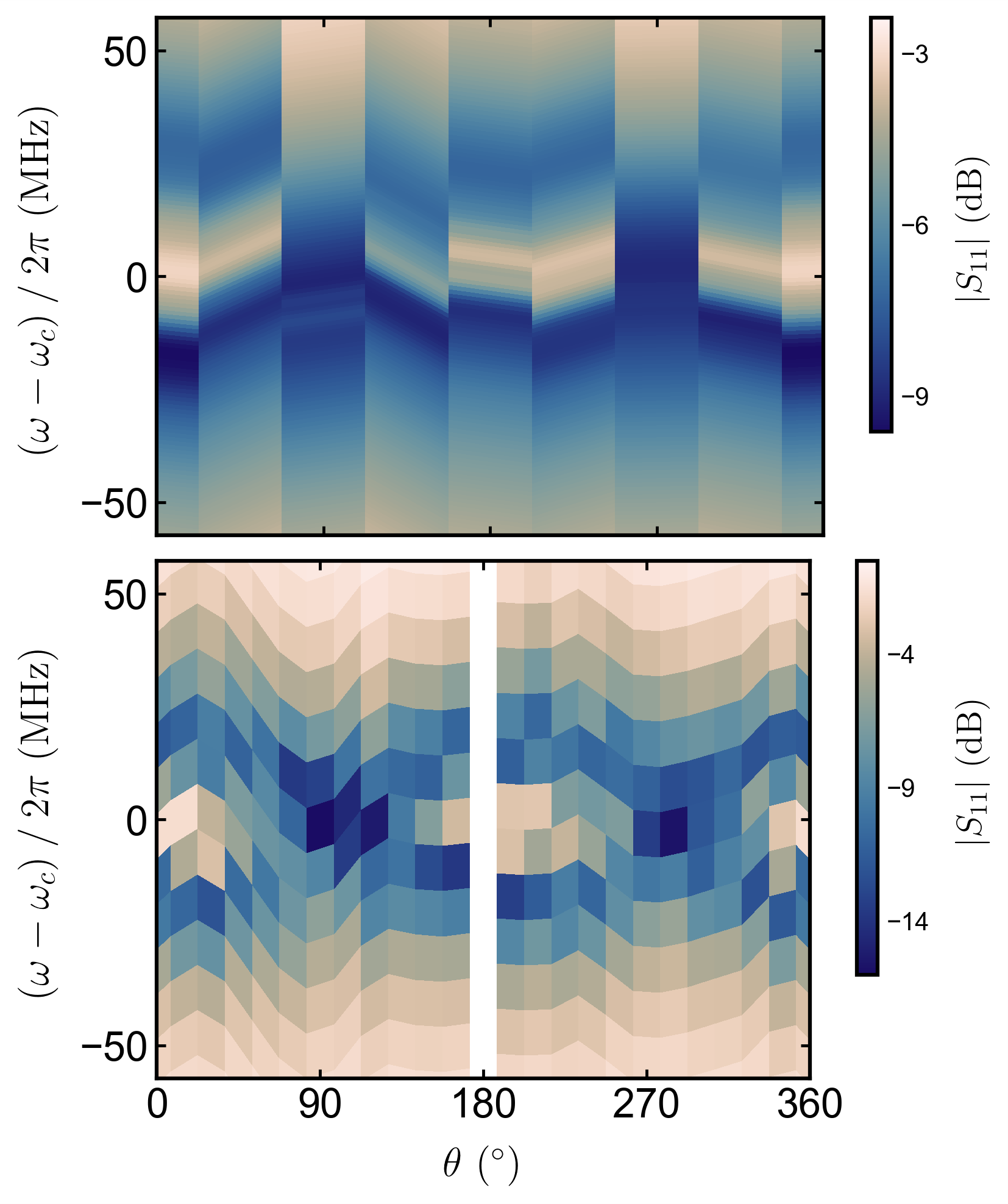}
  \caption{\textbf{Reflection spectrum at the magnon--cavity crossing as a function of the YIG position $\theta$.} (a) The measurement and (b) the COMSOL simulation. The vertical axis is the detuning from the cavity frequency of each angle. The grey dashed lines are the hybrid modes $\omega_c \pm g(\theta)$ from perturbation theory.}
  \label{fig:crossing_heatmaps}
\end{figure}

The coupling between the YIG sample and the ring resonator depends on where the sample sits around the ring because the microwave magnetic field it sees changes with position. To compare this dependence between the measurement and the simulation, we take, for every angle $\theta$, the reflection spectrum recorded exactly at the magnon--cavity crossing. At the crossing the bare magnon and cavity frequencies coincide, $\omega_0 = \omega_c$, and the two hybrid modes are separated by $2g$, so a single spectrum per angle carries the coupling strength $g(\theta)$. Figure~\ref{fig:crossing_heatmaps} stacks these spectra as a function of $\theta$.

In both the measurement and the simulation the two dips are furthest apart away from $\theta = 90^\circ$ and $270^\circ$ and merge into a single dip at the cavity frequency close to these angles. The perturbation-theory branches show the same behaviour, closing at $90^\circ$ and $270^\circ$ and opening between them. 

%%%%%%%%%%%%%%%%%%%%%%%%%%%%%%%%%%%%
\cleardoublepage
\section*{Supplementary Note C: Coupled Resonator Model}
\label{app:mp}
%%%%%%%%%%%%%%%%%%%%%%%%%%%%%%%%%%%%%

We consider two nominally identical copies of the ring resonator, with an inductive coupling between them, and the magnon coupled to resonator two. The Hamiltonian for the system is ($\hbar=1$)
\begin{align}
	\mathcal{H}_{sys} &= \sum_{n=1,2} \biggr[
	\widetilde{\omega}_a a_{+,n}^{\dag} a_{+,n} 
    + \widetilde{\omega}_a a_{-,n}^{\dag} a_{-,n} + 
	\beta a_{+,n}^{\dag} a_{-,n} + \beta^* a_{-,n}^{\dag} a_{+,n}
    +i\beta_d (a_{+,n}^{\dag} a_{-,n} + a_{-,n}^{\dag} a_{+,n}) \biggr]
	\\ \nonumber
	&\qquad\qquad + \widetilde{\omega}_0 m^{\dag} m  
	+ \biggr[
	g_{+} a_{+,2}^{\dag} m + g_{-} a_{-,2}^{\dag} m + H.c\biggr]
	\\ \nonumber
	&\qquad\qquad + [J(d) (a_{-,1}^{\dag} a_{-,2} + a_{+,1}^{\dag} a_{+,2}) + H.c.] 
	-iJ_d(d) (a_{-,1}^{\dag} a_{-,2} + a_{+,1}^{\dag} a_{+,2} + H.c.)
\end{align}
The first line describes a coherent and dissipative coupling between the CW ($+$) and CCW ($-$) modes of each resonator. The second line describes the coupling of resonator two to the magnon. In the final line, we include both a coherent and dissipative coupling between the CW modes of the different resonators, and similarly for the CCW modes. The cavity modes of the two resonators are coupled indirectly via the lossy electrons in the ring resonators; the dissipative component of the coupling accounts for the electronic losses. 

In the standing-wave basis,
\begin{align}
	a_{sn} = \frac{1}{\sqrt{2}}(a_{+,n}+a_{-,n}) \qquad\qquad 
	a_{cn} = \frac{1}{\sqrt{2}}(a_{+,n}-a_{-,n}) 
\end{align}
one finds
\begin{align}
	\mathcal{H}_{sys} &= \sum_{n=1,2} \biggr[\widetilde{\omega}_{s} a_{sn}^{\dag} a_{sn} 
	+ \widetilde{\omega}_{c} a_{cn}^{\dag} a_{cn} 
	+ g_{cs} (a_{sn}^{\dag} a_{cn} - a_{cn}^{\dag} a_{sn}) \biggr]  
	\\ \nonumber &\qquad\qquad  + \widetilde{\omega}_0 m^{\dag} m
	+ \biggr[g_{s} a_{s2}^{\dag} m + g_{c} a_{c2}^{\dag} m + H.c\biggr]
	\\ \nonumber
	&\qquad\qquad + [J(d) (a_{s1}^{\dag} a_{s2} + a_{c1}^{\dag} a_{c2}) + H.c.] 
	-iJ_d(d) (a_{s1}^{\dag} a_{s2} + a_{c1}^{\dag} a_{c2} + H.c.), 
\end{align}
where $\widetilde{\omega}_{c}$, $\widetilde{\omega}_{s}$, and $g_{cs}$ are given in Eq. (\ref{eq:params}), and the magnon-photon couplings are given in Eq. (\ref{eq:gcgs}). One may use input-output theory to determine the coupled equations of motion, as well as the $S$-matrix, and tune the coherent ($J(d)$) and dissipative ($J_d(d)$) couplings between the ring resonators to achieve good agreement between the spectra determined by the model and experimental results.

We note that the CW mode of resonator one was chosen to be coupled to the CW mode of resonator two, and similarly for the CCW modes. This choice was made because in the standing wave basis one finds the sine and cosine patterns of the different resonators are coupled to one another with the same phase, $J(d)(a_{s1}^{\dag} a_{s2} + a_{c1}^{\dag} a_{c2}) + H.c.$ With an inter-resonator coupling between the CW and CCW modes one obtains a relative phase shift between the sine and cosine patterns $J(d)(a_{s1}^{\dag} a_{s2} - a_{c1}^{\dag} a_{c2}) + H.c.$ In reality, the inter-resonator coupling will involve both capacitive and inductive components, and the form this coupling should take in the toy model is not clear. Further exploration of this detail is of interest, but lies beyond the scope of this paper.

One must use the full Hamiltonian to model the experimental data; however, with the YIG sphere placed at $\theta=0$, in the absence of backscattering, and with resonator ports included at $\theta=\pm \theta/2$, the sine modes decouple from the rest of the system and one may write down a simplified three mode model that captures the wireless readout and control of the magnon. We assume resonator one includes an input-output port at $\theta=-\pi/2$ and resonator two, which couples to the magnon, includes a port at $\theta=\pi/2$. The system-bath interaction is then
\begin{align}
	\mathcal{H}_{int} &= 
	i \int_{-\infty}^{\infty} d\omega\  
	[\kappa_{c1}^* b_1^{\dag}(\omega) a_{c1} - \kappa_{c1} b_1(\omega) a_{c1}^{\dag}]	
	+ i \int_{-\infty}^{\infty} d\omega\  
	[\kappa_{c2}^* b_2^{\dag}(\omega) a_{c2} - \kappa_{c2} b_2(\omega) a_{c2}^{\dag}]
    \\ \nonumber
	&= i \int_{-\infty}^{\infty} d\omega \ \boldsymbol{b}^{\dag} K^{\dag} \boldsymbol{a} 
	- i \int_{-\infty}^{\infty} d\omega \  \boldsymbol{a}^{\dag} K \boldsymbol{b},
\end{align}
where $\boldsymbol{b}^{\dag} = \begin{bmatrix} b_1^{\dag}  & b_2^{\dag}  \end{bmatrix}$, $\boldsymbol{a}^{\dag} = \begin{bmatrix} a_{c1}^{\dag}  & a_{c2}^{\dag} \end{bmatrix}$, and
\begin{align}
	K^{\dag} = \begin{bmatrix}
		\kappa_{c1}^* & 0  \\ 0 & \kappa_{c2}^*   \\
	\end{bmatrix} = \begin{bmatrix}
	\sqrt{2}\kappa & 0  \\ 	 
	0 & -\sqrt{2}\kappa   \\
	\end{bmatrix}.
\end{align}
Port 1 drives the $a_{c1}$ mode and port 2 drives the $a_{c2}$ mode. With the YIG sphere at $\theta=0$, one finds $g_s=-g_0\sin \theta = 0$ and $g_c=-ig_0\cos \theta = -ig_0$. For this arrangement, in the absence of backscattering, the sine and cosine modes are decoupled from one another. As only the cosine modes are coupled to the resonator ports, one may drop the sine wave modes from the system Hamiltonian to obtain
\begin{align}
	\mathcal{H}_{sys} &=  \widetilde{\omega}_{c} a_{c1}^{\dag} a_{c1} 
    + \widetilde{\omega}_{c} a_{c2}^{\dag} a_{c2}   
	+ \widetilde{\omega}_0 m^{\dag} m + \biggr[ g_{c} a_{c2}^{\dag} m + H.c\biggr]
	\\ \nonumber
	&\qquad\qquad + [J(d) a_{c1}^{\dag} a_{c2} + H.c.] 
	- iJ_d(d) [a_{c1}^{\dag} a_{c2} + H.c.], 
\end{align}
The coherent part of the coupling between $a_{c1}$ and $a_{c2}$ splits the modes. The dissipative part leads to asymmetry in their linewidths and the amplitudes of the transmitted/reflected signals. The dissipative component of the coupling has a significant effect on the experimental spectra, but it does not alter the wireless readout and control of the magnon, so we will drop it from subsequent consideration.

Following the input-output procedure, one finds the equations of motion for the system operators to be
\begin{align}
    \dot{a}_{c1} &= -iz_c a_{c1} - iJ(d) a_{c2} - \sqrt{2 \kappa_{\rm ext}} b_{1,in}
    \\ \nonumber
    \dot{a}_{c2} &= -iz_c a_{c2} -ig_c m - iJ^*(d) a_{c1} + \sqrt{2 \kappa_{\rm ext}} b_{2,in}
    \\ \nonumber
    \dot{m} &= -iz_m m -i g_c^* m 
\end{align}
where, as usual, $z_c$ now includes damping from the resonator ports and $\kappa_{\rm ext}=2\pi\kappa^2$. In frequency space, after eliminating the magnon from the equations of motion, one finds
\begin{align}
	i \Omega \boldsymbol{a} = \sqrt{2\pi} K \boldsymbol{b}_{in},
\end{align}
where the time evolution matrix is
\begin{align}
    \Omega = \begin{bmatrix} D_c^{-1} & -J(d) \\ 
                -J^*(d) & D_{cm}^{-1} \end{bmatrix}.
\end{align}
As above, the cavity mode response functions are
\begin{align}
	D_c = \frac{1}{\omega-z_c} \qquad\text{and}\qquad
	D_{cm} = \frac{1}{\omega-z_c-g_0^2 \chi_m}, 
\end{align}
where the magnon response is $\chi_m^{-1}=\omega-z_m$. The modes of the system follow from the zeros of 
\begin{align}
	\text{det}[\Omega] = D_{cm}^{-1} D_c^{-1} - |J_c|^2 = 
	\frac{(\omega-z_c)^2(\omega-z_m)-g_0^2(\omega-z_c)-|J(d)|^2(\omega-z_m)}{\omega-z_m}	
\end{align}
The susceptibility of the cavity modes is given by $\boldsymbol{a} = \chi K \boldsymbol{b}_{in}$, where the susceptibility tensor is
\begin{align}
	\chi= -i\sqrt{2\pi} \Omega^{-1} = \frac{-i \sqrt{2\pi}}{\text{det}[\Omega]}
	\begin{bmatrix} D_{cm}^{-1} & J(d) \\ J^*(d) & D_c^{-1} \end{bmatrix},
\end{align}
and the resulting $S$-matrix is
\begin{align}
	S = \mathbb{I} + \sqrt{2\pi} K^{\dag} \chi K. 
\end{align}
Having obtained the $S$-matrix, we consider transmission and reflection through the system.

As in Ref \cite{smith_exceptional_2026}, we take the coupling between resonators to be real. In this case the transmission is reciprocal
\begin{align}
	S_{12} = S_{21} 
	= \frac{i 2 \kappa_{\rm ext} J(d)}{D_{cm}^{-1} D_c^{-1} - J^2(d)}
    =\frac{i 2 \kappa_{\rm ext} J(d)(\omega-z_m)}
    {[(\omega-z_m)(\omega-z_c)-g_0^2](\omega-z_c)-J^2(d)(\omega-z_m)}.
\end{align}
With a complex inter-resonator coupling, transmission through the resonator complex will be non-reciprocal. Although of interest, non-reciprocity in the system under consideration lies beyond the scope of this paper. Naturally, transmission requires the coupling between resonators to be non-zero, $J(d) \neq 0$. The poles of the transmission function are the modes of the system, and its zero at the magnon-mode frequency, $z_m$, is an antimode. 

The reflection matrix elements are
\begin{align}
	S_{11} &=  
	1-\frac{i 2 \kappa_{\rm ext} D_{cm}^{-1}}{D_{cm}^{-1}D_c^{-1}-J^2(d)}
	= \frac{D_{cm}^{-1}(D_c^{-1}-i2 \kappa_{\rm ext})-J^2(d)}{D_{cm}^{-1}D_c^{-1}-J^2(d)}
    \\ \nonumber
    S_{22} &= 
    1-\frac{i 2 \kappa_{\rm ext} D_c^{-1}}{D_{cm}^{-1}D_c^{-1}-J^2(d)} 
	= \frac{(D_{cm}^{-1}-i2 \kappa_{\rm ext})D_c^{-1}-J^2(d)}{D_{cm}^{-1}D_c^{-1}-J^2(d)}
\end{align}
Let's unpack these. For $S_{11}$ we have
\begin{align}
	S_{11} = 1-\frac{i 2 \kappa_{\rm ext}}{D_c^{-1}-J^2(d) D_{cm}} = 
	 		1-\frac{i2 \kappa_{\rm ext}}{\omega-z_c-\frac{J^2(d)}
            {\omega-z_c-\frac{g_0^2}{\omega-z_m}}}
\end{align}
Here the empty resonator is driven. It couples to the loaded resonator which is coupled to the magnon. For $S_{22}$ we have
\begin{align}
	S_{22} = 1-\frac{i 2 \kappa_{\rm ext}}{D_{cm}^{-1}-J^2(d) D_c} 
	= 1-\frac{i 2 \kappa_{\rm ext}}
    {\omega-z_c-\frac{g_0^2}{\omega-z_m}-\frac{J^2(d)}{\omega-z_c}}.
\end{align}
We see the driven cavity mode is coupled to both the magnon and to the second, degenerate resonator. A comparison between the toy model and the experimental spectra is provided in Supplementary Note D.

%%%%%%%%%%%%%%%%%%%%%%%%%%%%%%%%%%%%%%%%%%%%%%%%%%%%%%%%%%%%%%%%%%%%%%%%%%%%%%%%%%%%%%%%%%%%%
\section*{Finite-Element Validation of the Inter-Ring Coupling}
%%%%%%%%%%%%%%%%%%%%%%%%%%%%%%%%%%%%%%%%%%%%%%%%%%%%%%%%%%%%%%%%%%%%%%%%%%%%%%%%%%%%%%%%%%%%%

\begin{figure}[h]
\centering
\includegraphics[width=0.7\linewidth]{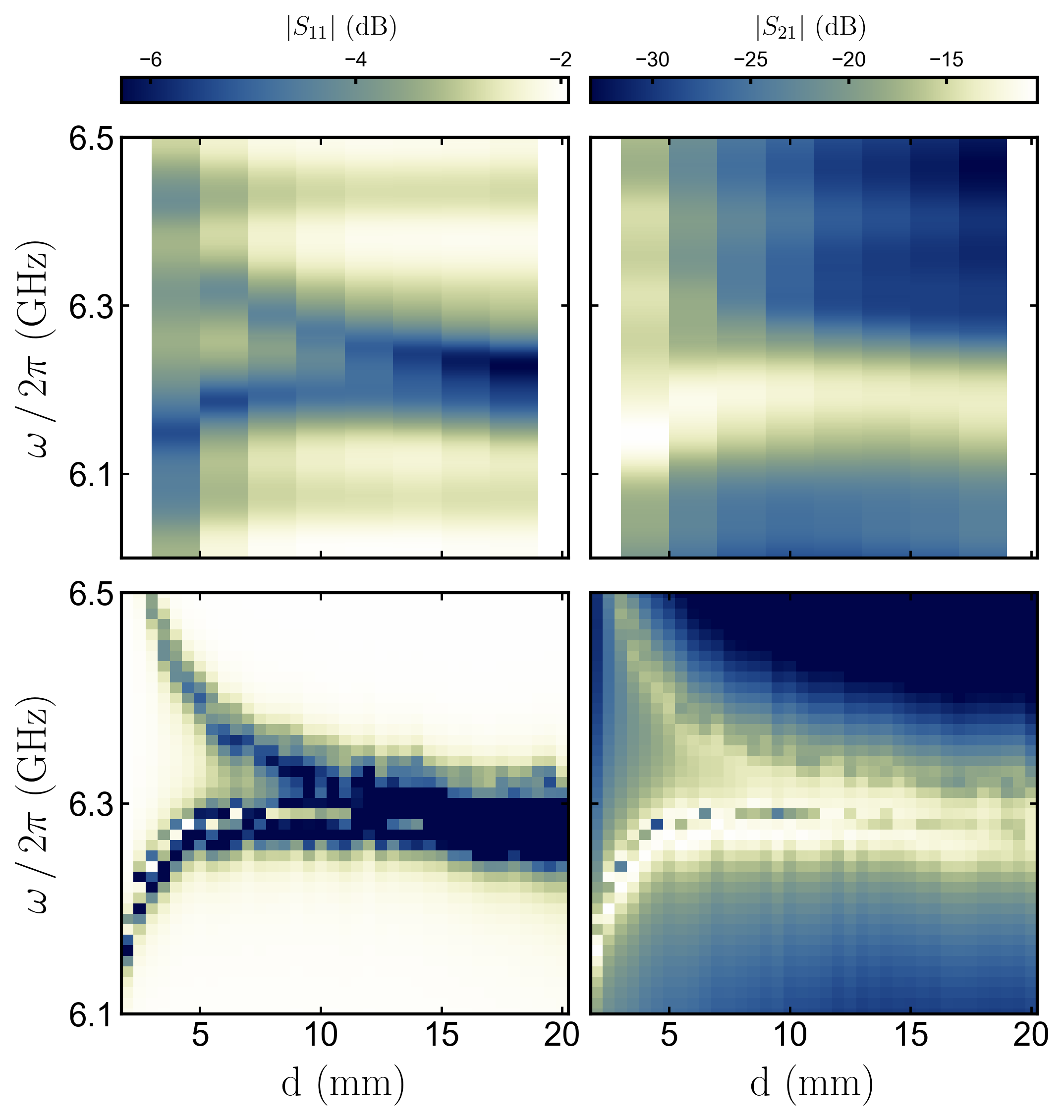}
\caption{\textbf{Measured versus simulated coupling spectra.} Amplitude of $|S_{11}|$ (a,c) and $|S_{21}|$ (b,d) versus separation $d$ and frequency, for the measurement (a,b, averaged over the magnetic-field sweep) and the finite-element simulation (c,d).}
\label{fig:sup_fig_measVsSim}
\end{figure}

Fig.~\ref{fig:fig2}(c) of the main text extracts the inter-ring coupling $J(d)$ from the mode splitting in $S_{11}$; beyond $d \approx 10$~mm this splitting approaches the resolution limit set by the linewidth, and $J$ is instead read off from dip/peak positions across all four measured channels. As an independent check of this procedure, Fig.~\ref{fig:sup_fig_measVsSim} compares the measured amplitude response against a finite-element simulation (COMSOL) of the same stacked-ring geometry, swept over $d$ and frequency with the magnon far detuned.

\begin{figure}[ht]
\centering
\includegraphics[width=0.55\linewidth]{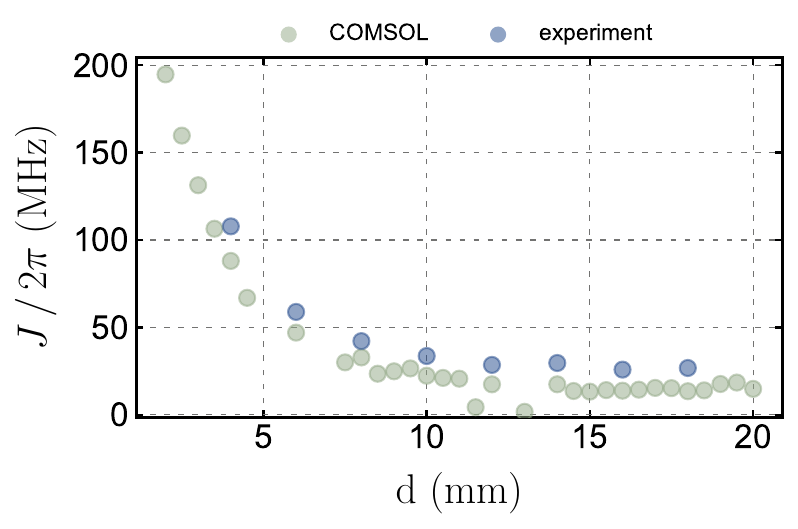}
\caption{\textbf{Inter-ring coupling: measurement versus finite-element simulation.} $J(d)/2\pi$ extracted from the measured mode splitting (as in Fig.~\ref{fig:fig2}(c) of the main text) and from the COMSOL-simulated $S_{11}$ dip splitting, over the same range of separations $d$.}
\label{fig:sup_fig_Jcompare}
\end{figure}

The simulation reproduces the measured structure in both channels: the two mode branches and their convergence with increasing $d$. The COMSOL splitting is extracted by fitting a double-Lorentzian curve to the simulated $S_{11}$ dip, $J_{\rm sim}$ then follows from half the resulting splitting, the bare detuning between the identical simulated rings being negligible by construction (as shown in Fig.~\ref{fig:sup_fig_Jcompare}). The simulated and measured $J(d)$ show excellent agreement across the full range and decay with the same functional form, corroborating the method used to extract $J(d)$ from the measurement in Fig.~\ref{fig:fig2}(c) of the main text, including beyond $d \approx 10$~mm where the splitting is only marginally resolved. Similar results for the splitting of the coupled resonator modes can be obtained from the coupled mode model. 

  \cleardoublepage
%%%%%%%%%%%%%%%%%%%%%%%%%%%%%%%%%%%%%%%%%%%%%%%%%%%%%%%%%%%%%%%%%%%%%%%%%%%%%%%%%%%%%  
\section*{Supplementary Note D: Full Experimental $S$-parameter set for the magnon-mediated transmission and comparison with the toy model}
\begin{figure}[h]
\centering
\includegraphics[width=0.925\linewidth]{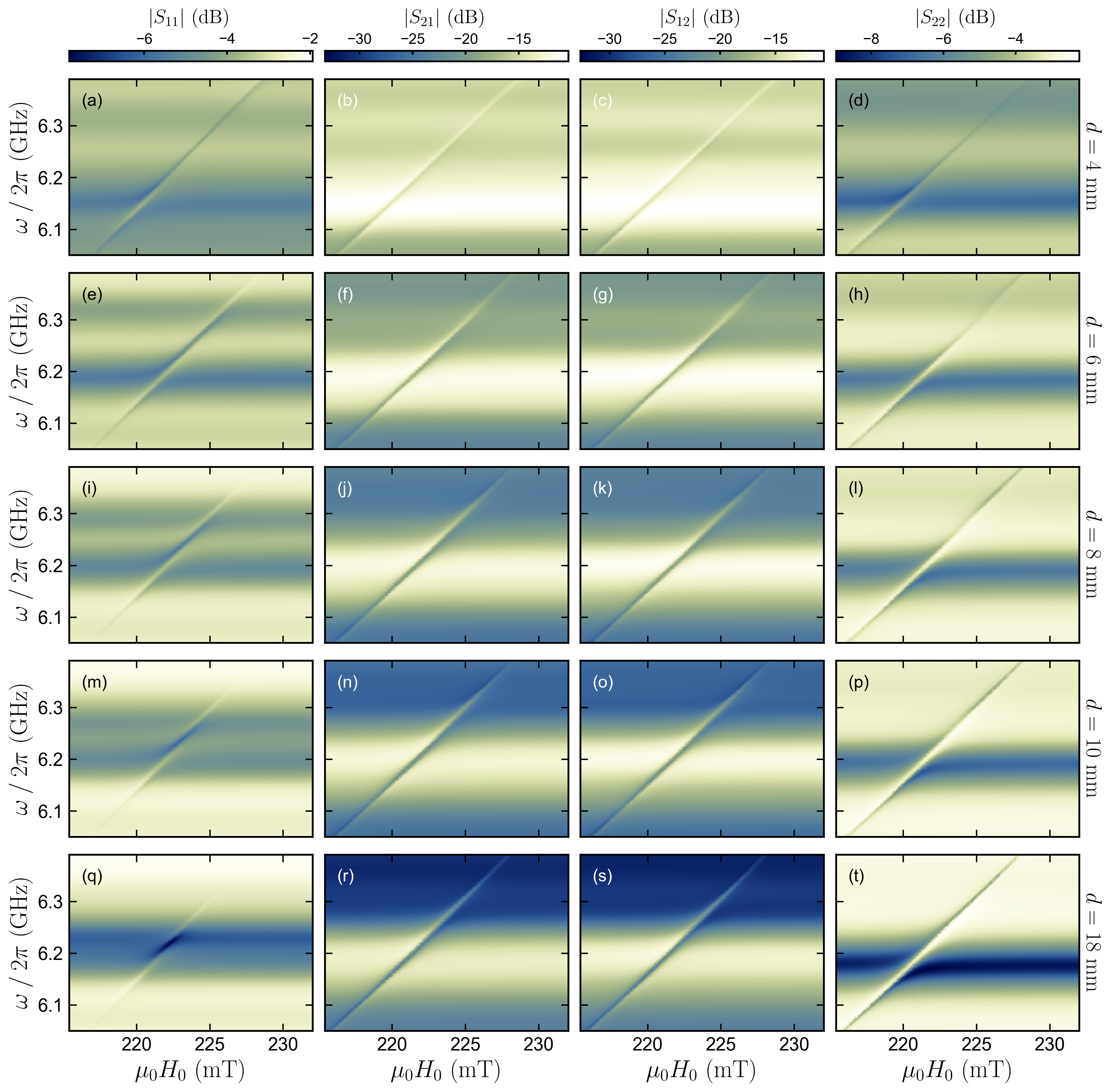}
\caption{\textbf{Full experimental $S$-parameter set behind Fig.~\ref{fig:fig3}.} Field-swept spectra with the YIG sphere at $\theta = 0$ on ring~2, for separations $d = 4$~mm (a)--(d), $6$~mm (e)--(h), $10$~mm (i)--(l) and $18$~mm (m)--(p). Columns, left to right: $S_{11}$ (reflection, port~1, no sphere), $S_{21}$ and $S_{12}$ (transmission between the rings), $S_{22}$ (reflection, port~2, the sphere's own ring).}
\label{fig:sup_fig3_all4}
\end{figure}

%%%%%%%%%%%%%%%%%%%%%%%%%%%%%%%%%%%%%%%%%%%%%%%%%%%%%%%%%%%%%%%%%%%%%%%%%%%%%%%%%%%%%
Fig.~\ref{fig:sup_fig3_all4} extends Fig.~\ref{fig:fig3} of the main text to the complete set of measurements: reflection and transmission at both ports, $S_{11}$, $S_{21}$, $S_{12}$ and $S_{22}$, over the separations $d = 4$, 6, 10 and 18~mm.

% The reflection columns ($S_{11}$, $S_{22}$) show the magnon crossing each ring's share of the hybridised modes. The crossing is present in both channels at every separation, including in $S_{11}$, the reflection of the ring that carries no sphere -- consistent with the hybrid, shared character of the modes discussed in the main text. The avoided crossing in reflection has the same size at every separation shown, reflecting the fixed magnon--photon coupling set by the sphere's position on its own ring.

% The transmission columns ($S_{21}$, $S_{12}$) carry the same crossing across the gap at every separation. The two transmission channels agree with each other, as expected for a reciprocal system, and both show the smaller splitting relative to reflection quoted in the main text. As the separation increases, the two ring modes -- whose splitting is set by the inter-ring coupling $J$, not by the magnon -- move closer together, and by $d = 18$~mm they are no longer individually resolved in any of the four channels; the magnon then interacts with the merged cavity response as a single feature, as described in the main text.

In order to validate the coupled mode model, we use it here to reproduce the experimental spectra shown in Fig. \ref{fig:sup_fig3_all4}. We consider $\omega_a=6.21$ GHz resonators with the magnon at $\theta=0$ in which case the magnon-photon coupling between resonator two and the cosine mode is $g_c=-ig_0$ and $g_s=0$. The strength of the magnon-photon coupling is set to the experimentally relevant value of $g_0=10$ MHz. For each resonator, we include the backscattering parameters from Supplementary Note A: Extraction of the backscattering. That is, we set $|\beta|/2\pi=17.75$ MHz and $\phi_{\beta} = 197^\circ$, with a dissipative component of $\beta_d/2\pi = 5$ MHz. Recall this reproduces the offset of the two peaks seen in the reflection spectra, $S_{11}$ and $S_{22}$. The damping of the magnon mode is set to $\kappa_m/2\pi=3$ MHz. In the upper panels of Fig. \ref{fig:ERthetaJd} we set the inter-resonator coupling to $J=100$ MHz corresponding to a ring separation of $d \approx 4$ mm. In order to capture the asymmetry of the hybridised cavity modes, the dissipative component of the inter-resonator coupling is set to $J_d=15$ MHz. The middle panels show the calculated spectra with $J=45$ MHz and $J_d=3$ MHz corresponding to $d \approx 8$ mm, and the lower panels show the spectra with $J=20$ MHz and $J_d=1$ MHz corresponding to $d \approx 18$ mm.

Although discrepancies exist between the calculated and measured transmission spectra, over all, we find their agreement to be good. The modes are split by $J$ and the dissipative component of the inter-resonator coupling leads to asymmetry in their magnitudes. The dissipative component of the mode coupling due to backscattering leads to a stronger dip in the reflection spectrum in $S_{22}$ than in $S_{11}$, consistent with the experiment. In both the calculated spectra and the experimental data, there is a strong dip at the magnon mode frequency in $S_{11}$, and strong reflection in $S_{22}$, in between the horizontal dips seen in the reflection spectrum. In the experiment, one sees two dips in the reflection spectrum for $d=4 \rightarrow 10$ mm in $S_{11}$ corresponding to transmission by the hybridised cavity modes, but the high frequency reflection dip appears absent in $S_{22}$, contrary to the calculation, and the reflection dips are more prominent in the experiment than in the calculation. One may be able to account for this by making further improvements to the model, for example, by accounting for the detuning of the two cavity modes, and through refinements of the system parameters, in particular, accounting for their variation as the resonators are separated, but this lies outside the scope of the present report.

\begin{figure}[h]
\centering 
\hspace*{-1.1cm}
\includegraphics[width=1.15\linewidth]{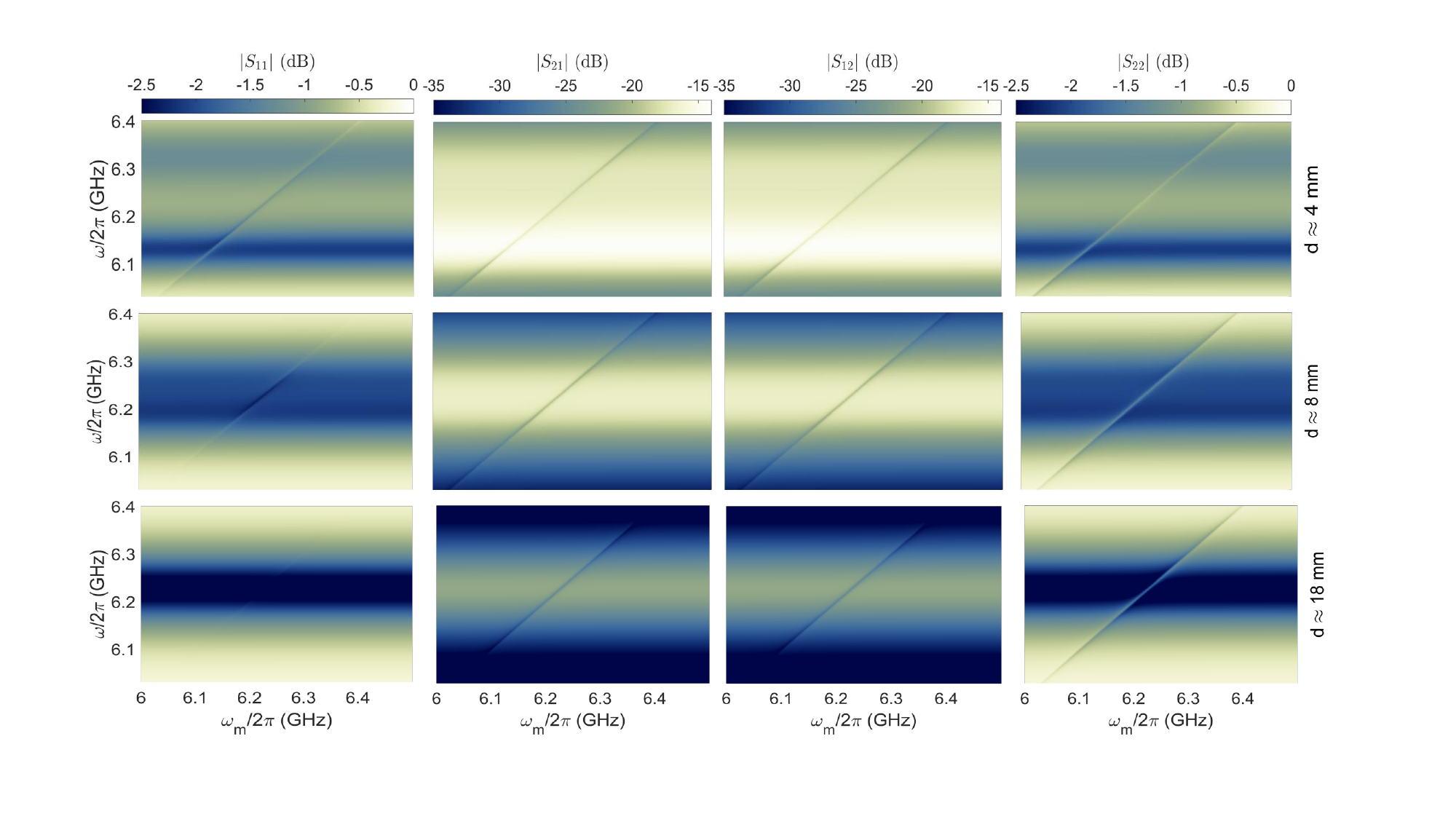}
\caption{\textbf{Full theoretical $S$-parameter.} Transmission and reflection spectra calculated with the coupled mode model including the effects of backscattering and dissipative couplings. The two resonators are assumed to be identical. We use experimentally relevant parameters corresponding resonator separations of $d \approx 4$ mm, $d\approx 10$ mm, and $d \approx 18$ mm, and tune the dissipative component of the inter-resonator coupling to approximate the experimental data. }
	\label{fig:ERthetaJd}
\end{figure}

\cleardoublepage

\end{document}